\documentclass[pra,aps,twocolumn,showpacs]{revtex4}
\usepackage{graphicx}
\usepackage{amsmath}
\usepackage{bm} 

\usepackage{times} 

\newcommand{\ie}{{\it i.e.\ }}
\newcommand{\eg}{e.g.\ }
\newcommand{\cf}{cf.\ }
\newcommand{\dpSF}{\Delta p^\mathrm{SF}}
\newcommand{\dpRSF}{\Delta p^\mathrm{RSF}}
\def\ket#1{\left|{#1}\right>}

\def\vecb#1{{\bm{#1}}}

\begin{document}

\title{A Spectroscopic Method to Measure the Superfluid Fraction of an Ultracold Atomic Gas}

\author{S.~T.~John, Z.~Hadzibabic and N.~R.~Cooper}
\affiliation{Cavendish Laboratory, University of Cambridge, J.~J.~Thomson Ave., Cambridge CB3~0HE, U.K.}

\begin{abstract}

  We perform detailed analytical and numerical studies of a recently
  proposed method for a spectroscopic measurement of the superfluid
  fraction of an ultracold atomic gas~[N. R. Cooper and Z. Hadzibabic,
  Phys.\ Rev.\ Lett.\ \textbf{104}, 030401 (2010)].  Previous
  theoretical work is extended by explicitly including the effects of
  non-zero temperature and interactions, and assessing the
  quantitative accuracy of the proposed measurement for a one-component Bose gas. We show that for
  suitably chosen experimental parameters the method yields an
  experimentally detectable signal and a sufficiently accurate
  measurement. This is illustrated by explicitly considering two key
  examples: First, for a weakly interacting three-dimensional Bose gas
  it reproduces the expected result that below the critical
  temperature the superfluid fraction
  closely follows the condensate fraction. Second, it allows a clear quantitative differentiation
  of the superfluid and the condensate density in a strongly
  interacting Bose gas.

\end{abstract}
\date{November 10, 2010}


\maketitle

\section{Introduction}

Ultracold atomic gases provide a highly controllable ex\-per\-i\-men\-tal
setting for studies of many-body quantum phenomena such as
Bose-Einstein condensation~\cite{Anderson1995} and
superfluidity~\cite{ColdgasSF1,ColdgasSF2,ColdgasSF3} (for a review see~\cite{RevModPhys2008}).  
The physical phenomena studied in these systems are often
analogous to those occurring in other many-body systems, in particular
solid state materials and liquid helium. Moreover, the flexibility in
experimentally designing the geometry and interactions in atomic gases
offers the possibility to experimentally access physical regimes which
are of theoretical interest but could so far not be observed.  At the
same time, the experimental probes used in atomic physics are often
quite different from those traditionally used in condensed matter
experiments and discussed in the theoretical literature. In
particular, for atomic gases there is no general experimentally
established method allowing a quantitative measurement of the
superfluid fraction, which is traditionally defined through the
fluid's response to rotation and the emergence of a non-classical
moment of inertia~\cite{Leggett1999}. In classic experiments on liquid
helium~\cite{Andronikashvili1946}, under rotation of the walls of the
container a perfect superfluid remains metastable in the zero angular
momentum state as long as the rotation rate does not exceed the
critical velocity for a superfluid flow. More generally, the fraction
of the fluid which does not rotate with the walls quantitatively
defines the superfluid fraction.

Only very recently some ideas on how to measure the superfluid
fraction of an ultracold atomic gas have been
formulated~\cite{Riedl2009,HoZhou2009,CooperHadzibabic2010}. Specifically,
in Ref.~\onlinecite{CooperHadzibabic2010} a spectroscopic method was proposed,
which closely follows the traditional definition of the superfluid
density but allows a signal readout which plays to the strengths of
atomic physics. This proposal builds on the recent developments in
the use of optical fields to induce artificial gauge fields for ultracold
atoms~\cite{dalibardreview}. The key idea is that if slow rotation of
the gas is induced by an optical gauge field, this creates a natural
coupling between the external (angular momentum) and the internal
(hyperfine) atomic degrees of freedom.  Measuring the populations of
hyperfine states in an atomic cloud then allows a direct readout of
the angular momentum induced by the rotation, and thus a measurement
of the moment of inertia and the superfluid fraction.

The basic connection between the hyperfine populations and the
superfluid fraction was pointed out in Ref.~\onlinecite{CooperHadzibabic2010} by
considering the difference between a perfect superfluid with no
angular momentum and a fully relaxed gas which exhibits the classical
value of the moment of inertia.  Here we extend this theoretical work
in several ways. First, we include in our calculations the effects of
non-zero temperature and interactions in the gas, which broaden the
distribution of angular momenta around zero for a metastable
superfluid, and around the classical value set by the imposed rotation
for a fully relaxed gas. This allows us to assess the quantitative
accuracy of the proposed measurement, and to estimate the experimental
parameters which in practice would allow a good compromise between the
theoretical accuracy of the method and the experimentally relevant
size of the readout signal.  Second, we explicitly calculate the
expected experimental signal in two important cases: For a weakly
interacting Bose gas we show that the spectroscopically deduced
superfluid fraction closely follows the condensate fraction below the
critical temperature; this confirms that the proposed method gives the
expected result in this well understood limit. On the other hand, in
the limit of strong interactions the superfluid and the condensate
fraction of the gas can be quite different, as is known from the case
of liquid helium~\cite{Griffin1996}. We show that in this limit the spectroscopic
measurement is sufficiently accurate to allow a clear experimental
distinction between the two quantities.

The paper is organized as follows. In Section~\ref{sec:set-up} we
lay out the theoretical background on the concept of superfluid density
and its connection to hyperfine populations in an atomic cloud rotated
with use of optically induced gauge potentials. In
Section~\ref{sec:3-level-system} we give some more details on a specific
implementation of gauge fields in an atomic system. In
Section~\ref{sec:corrections} we discuss the quantitative theoretical
corrections to the mapping between the hyperfine populations and the
superfluid fraction. To illustrate the effect of these corrections we
first consider a normal gas at non-zero temperature; this already
allows us to anticipate suitable experimental parameters which lead to
sufficiently small theoretical inaccuracies and sufficiently large
experimental signals. In Section~\ref{sec:interactions} we extend our
calculations to interacting gases, at both zero and non-zero
temperature. Our results for the expected experimental signals in both
weakly and strongly interacting gases are presented in
Section~\ref{sec:results}. Finally, Section~\ref{sec:summary} contains
a summary of the paper.

\section{Superfluid density}
\label{sec:set-up}

The concept of a superfluid density, or superfluid fraction,
originates in the two-fluid model for the hydrodynamics of superfluid
$^4$He, proposed by Tisza~\cite{tisza1,tisza2} and Landau~\cite{Landau}.
The fluid, of total density $\rho$, is assumed to consist of a
superfluid component, of density $\rho_s$, which has vanishing
viscosity and flows without dissipation, and a normal component, of
density $\rho_n= \rho-\rho_s$.
Landau proposed~\cite{Landau} how to measure these separate
components. He envisaged taking superfluid helium at rest in its
container, and slowly rotating the walls at a constant angular
velocity $\omega$.  The normal component equilibrates and moves along with the
rotating walls; however, the superfluid component is unaffected and
remains at rest. Since only the normal component moves, the moment of
inertia of the fluid is determined by $\rho_n$, and its ratio to the
expected classical moment of inertia (defined by the total
density $\rho$) provides the normal fraction $\rho_n/\rho$ and hence
the superfluid fraction $1-\rho_n/\rho$.  Note that it is necessary
that the trap is \emph{not} perfectly rotationally symmetric (\ie the
walls must be rough), so that the normal fluid can relax into the
steadily rotating state and come into equilibrium by changing its
angular momentum.

This method was implemented for superfluid helium in the classic
experiments by Andronikashvili~\cite{Andronikashvili1946}. In those
experiments it was not the container that was rotated, but a stack of
disks embedded in the fluid. Still, the disks drag just the normal
fluid, so measurements of the moment of inertia of the disks (using a
torsional oscillator) allowed a determination of the normal and
superfluid fractions.

\begin{figure}
  \centering
  \includegraphics[scale=0.4]{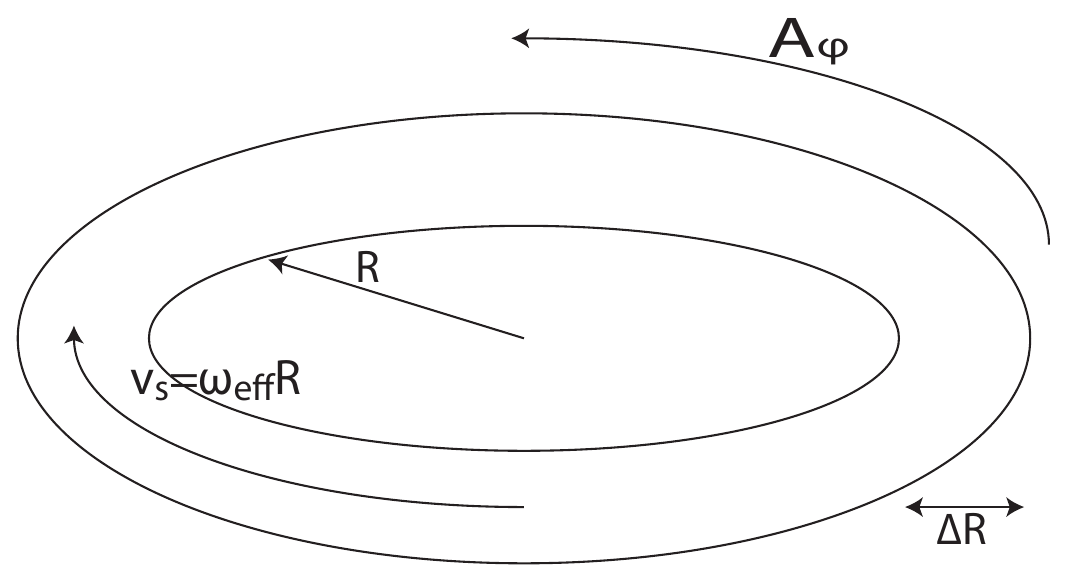}
  \caption{Geometry of the vessel considered in this paper: a torus of radius $R$ with transverse dimensions $\Delta R \ll R$. $A_\varphi$ is the gauge potential introduced by light fields, corresponding to rotation with angular frequency $\omega_\textrm{eff}$. This induces a superfluid flow with speed $v_s$ in the direction opposite to $A_\varphi$.}
  \label{fig:sketch1}
\end{figure}

The non-classical moment of inertia arising from the superfluid
component provides the standard definition of the superfluid
fraction~\cite{Leggett1999}.  To discuss this theoretically, it is
customary to consider the fluid to be contained in a ring-shaped
toroidal vessel with a radius $R$ much larger than its transverse
dimensions $\Delta R$, \cf Fig.~\ref{fig:sketch1}.
In this case, the classical moment of inertia for $N$ atoms of mass
$M$ is given by $I_\textrm{cl} = NMR^2$.  We shall assume this
geometry throughout this paper --- in part for theoretical simplicity,
but also for practical reasons discussed further below.
The superfluid fraction is then defined~\cite{Leggett1999} by the average angular
momentum $\langle L\rangle$ picked up under rotation of the vessel
with an angular frequency $\omega$:
\begin{equation}
  \label{eq:superfluid-definition}
  \frac{\rho_s}{\rho} \equiv 1 - \lim_{\omega\to 0} \left(\frac{\langle L\rangle}{I_\textrm{cl} \omega}\right) .
\end{equation}
The limit $\omega\to 0$ of slow rotation of the vessel is required
such that the velocity of the walls of the container, $\omega R$, does
not exceed the critical velocity of the superfluid, $\omega R <
v_{\rm crit}$. Since finite-size effects can be much more important in
ultracold atomic gases than in typical experiments on superfluid
helium, it is helpful to expand further on this point. For a very small
rotation rate, namely when $\omega < \hbar/2MR^2$, even an {\it
  ideal} Bose gas shows a non-classical moment of inertia, and could thus be
considered to be superfluid~\cite{blatt1955}.  (For $\omega <
\hbar/2MR^2$, the lowest-energy single particle state remains the state
with vanishing angular momentum, so an ideal condensate has $L=0$.)
We define superfluidity in the strongest sense of
Ref.~\onlinecite{blatt1955}, which is the sense that is conventionally
applied: the angular momentum is measured with an imposed rotation
frequency in the range $ \hbar/2MR^2 < \omega <
v_{\rm crit}/R$. The lower limit excludes the ideal Bose gas as a
superfluid. For a large system, $R\gg \hbar/Mv_{\rm crit}$, this constitutes a wide range of
frequencies. For a weakly interacting Bose gas, $v_{\rm crit} \sim
\hbar/M\xi$ (where $\xi$ is the healing length, which will be further considered in Section~\ref{sec:interactions}), and so the range of
frequencies is wide for $R \gg\xi$.

This method could, in principle, be applied to ultracold atomic gases,
using a rotating deformation of the trap to re\-pre\-sent the rotating
walls of the container.  However, in practice this is difficult, as a
measurement of the induced angular momentum or mass flow is difficult
for ultracold gases.  For harmonically trapped gases, an ingenious method of
measuring the angular momentum has been applied~\cite{zambelli1998,chevy2000,Riedl2009}.  A
theoretical proposal has shown how the superfluid density could be
extracted more generally if local imaging is
possible~\cite{HoZhou2009}.
In this paper we explore an alternative theoretical
proposal~\cite{CooperHadzibabic2010} in which the rotation is simulated
by an optically induced gauge field.  A key feature of this method is
that it allows one to  measure the angular
momentum \emph{spectroscopically} and hence deduce the superfluid fraction of an ultracold atomic
gas.

The basis of the idea is the coupling of light with orbital angular
momentum~\cite{LaserPhotonOAM} to internal atomic spin states, thereby
creating an azimuthal gauge field $A_\varphi$~\cite{juzeliunasoam}. The azimuthal
gauge leads to the same effects as rotation, albeit in a slightly
different manner.  In the presence of the gauge field one must make a
distinction between the \emph{canonical} momentum $\vecb{p}_{\rm can}$ and the \emph{kinetic} momentum $\vecb{p} = \vecb{p}_{\rm can} -
\vecb{A}$. In the absence of a gauge field, a superfluid that is at
rest in the toroidal vessel has no winding number of its phase (no
vortices) and corresponds to the case of vanishing canonical momentum
$\vecb{p}_{\rm can}=0$.  This does not change when the gauge field is
introduced. However, as the gauge field is turned on, the superfluid
picks up a non-zero velocity $\vecb{p}/M = - \vecb{A}/M$. On the other
hand, for a normal fluid, as the gauge field is switched on, the fluid will
always stay at rest with the walls of the container (provided they are
rough).  Thus, compared to the rotating container discussed above,
here it is the superfluid that moves while the normal fluid stays at
rest.  The case of the gauge field is, in fact, exactly equivalent to
a rotating vessel, but where one views the system in the rotating
frame of reference and so experiences the trap (and normal fluid) to be at
rest.

To make this discussion more precise, note that when the optically
induced gauge field is on, the atoms experience an effective
dispersion relation which can be written as
\begin{equation}
  \label{eq:effective-dispersion}
  E \simeq E_0 + \frac{\hbar^2}{M^* R^2} \left(\frac{\ell^2}{2} - \ell\ell^*\right) + {\cal O}(\ell^3) ,
\end{equation}
where $\ell$ is the angular momentum in units of $\hbar$, such that 
it is quantized to integer values. $M^*$ is a
new effective mass of the atoms, and $\ell^*$ is the rotational shift
due to the gauge field.  (We will derive this in the next
section.)  This effective dispersion is equivalent to an azimuthal
gauge field $A_\varphi = \hbar\ell^* / R$.  It corresponds to being in
a frame of reference rotating with an effective angular velocity
\begin{equation}
  \omega_{\rm eff} = \frac{\hbar \ell^*}{M^* R^2} ,
\end{equation}
\eg a particle with $\ell=0$ will have an angular group velocity
$(1/\hbar)dE/d\ell = -\omega_\textrm{eff}$.  Following the above
discussion, if $\omega_{\rm eff}$ is
slowly increased from zero, the superfluid will remain in its
original state with $\langle L\rangle=0$ but will flow with speed
$v_s=\omega_\textrm{eff} R$ in the direction opposite to $A_\varphi$.
In contrast, the normal fluid will pick up
an average angular momentum of $\hbar\ell^*$ per particle but will retain zero average velocity.

A key feature of the proposal~\cite{CooperHadzibabic2010} is that since the
gauge field is generated by mixing internal hyperfine states of the
atoms, this provides a natural coupling between internal and external
degrees of freedom.  Spectroscopically measuring the population of the
different internal spin states allows the average angular momentum
per atom $\hbar\langle\ell\rangle$ to be deduced. We will focus on the
population difference of the hyperfine states $\ket{+1}$ and
$\ket{-1}$ in a three-level system (with amplitudes $\psi_{+1}$ respectively $\psi_{-1}$), but this could be adapted to other
internal structures. (The three-level system will be described in
further detail in the next section.) For a single atom with angular
momentum $\ell$ we define the population difference as
\begin{equation}
 \Delta p_\ell \equiv \left| \psi_{-1}(\ell) \right|^2 - \left| \psi_{+1}(\ell) \right|^2 .
\end{equation}
A measurement of the populations $N_{+1}$ and $N_{-1}$ for a gas of such atoms
leads to the fractional population difference $\Delta p$, which may be expressed in terms of $\Delta p_\ell$ as
\begin{align}
\nonumber
  \Delta p \equiv \frac{N_{-1} - N_{+1}}{N}
  &= \frac{\sum_\ell \langle n_\ell\rangle \left[\left| \psi_{-1} \right|^2 - \left| \psi_{+1} \right|^2\right]}{\sum_\ell \langle n_\ell\rangle}\\
  &= \frac{\sum_\ell \langle n_\ell\rangle \Delta p_\ell}{\sum_\ell \langle n_\ell\rangle} .
\end{align}
Within the assumption that $\Delta p_\ell$ can be expanded to first order,
\begin{equation}
\Delta p_\ell \simeq   \Delta p_0 + \Delta p' \ell + {\cal O}(\ell^2) ,
  \label{eq:dp-single}
\end{equation}
one can deduce the angular momentum expectation value
\begin{equation}
  \label{eq:ell-measurement}
  \frac{\langle L\rangle}{N \hbar} \equiv \langle\ell\rangle \equiv \frac{\sum_\ell \langle n_\ell\rangle \ell}{\sum_\ell \langle n_\ell\rangle} 
  \simeq \frac{\Delta p - \Delta p_0}{\Delta p'} .
\end{equation}
Putting this back into Eq.~\eqref{eq:superfluid-definition}, one gets
\begin{equation}
  \label{eq:superfluid-measurement}
  \frac{\rho_s}{\rho} \simeq 1 - \lim_{\ell^*\to 0} \frac{\Delta p - \Delta p_0}{\Delta p' \ell^*} ,
\end{equation}
where the appropriate moment of inertia  $I_\textrm{cl} = N M^* R^2$ has been used.
For a perfect superfluid we would expect to find $\Delta p \equiv
\Delta p_0$, whereas for a normal fluid we would expect to find
$\Delta p \equiv \Delta p_0 + \Delta p' \ell^*$, thus allowing us to
distinguish between the two. 

The main goal of this paper will be to analyse the {\it quantitative}
accuracy of (\ref{eq:superfluid-measurement}), allowing for
corrections that arise from the higher-order terms that are neglected
in Eqs.~(\ref{eq:effective-dispersion},\ref{eq:dp-single}).  However,
for now, note that the spectroscopic technique should show a clear
{\it qualitative} signature of superfluidity. For a normal fluid (or
the normal fraction), it does not matter in which order one increases
$\ell^*$ and cools the gas to its final temperature:
that is, these two operations ``commute''.  However, for the
superfluid fraction these operations do not commute: if one first
cools at $\ell^*=0$ and then imposes non-zero $\ell^*$, the superfluid
is pushed into a (metastable) state in which it is moving with respect
to the walls of the container; if one first imposes non-zero $\ell^*$
and then cools, the superfluid will be formed at rest with respect
to the walls.  Thus, depending on the
order, the system is led either to the non-relaxed metastable
superfluid condensed in a state of vanishing (canonical) angular
momentum $\ell_c=0$, which we label ``SF'', or to the relaxed
superfluid, condensed in the ground state $\ell_c=\ell^*$, which
we label ``RSF''. The two cases will have different fractional
hyperfine populations, $\dpRSF\neq \dpSF$, so the change in population
allows a clear qualitative signature of metastable superfluid flow.

\section{Optically dressed states}
\label{sec:3-level-system}

There are several well-established theoretical proposals for how
optical fields can be used to create fictitious gauge fields in
neutral atomic gases~\cite{dalibardreview}. The scheme we follow here
is closely related to that implemented in the experiments of the NIST
group~\cite{Spielman2009,LinSpielman2009}. However, it is adapted to
generate an azimuthal vector potential, by using optical beams with
orbital angular momentum~\cite{juzeliunasoam}. As described above,
throughout this work we assume that the gas is confined in a toroidal
trap (Fig.~\ref{fig:sketch1}) with radius $R$ large compared to its
transverse dimensions $\Delta R$. This simplifies the experimental
implementation of the azimuthal gauge field, as it will be sufficient
to require that the optical fields are uniform only over the range
$\Delta R$.

\begin{figure}
  \centering
  \includegraphics[scale=0.7]{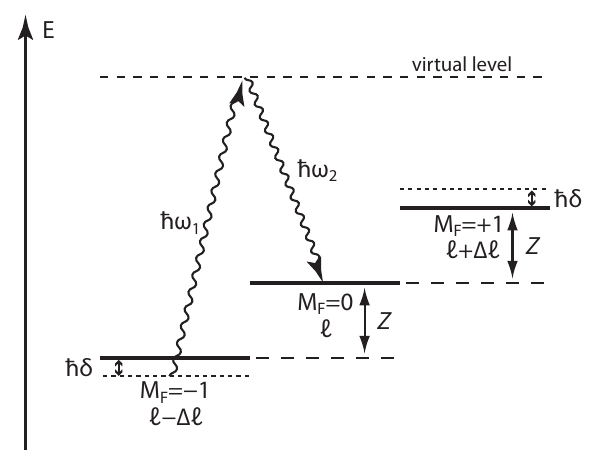}
  \caption{Sketch of a two-photon Raman transition between hyperfine levels, adapted from Spielman et~al.~\cite{Spielman2009,LinSpielman2009}.}
  \label{fig:raman}
\end{figure}

We consider atoms with three hyperfine levels~\cite{Spielman2009} in their
electronic ground state, \eg ${}^{23}$Na with $F=1$.  The degeneracy
of the three hyperfine states $M_{F}=0,\pm1$ is lifted by applying a
weak external magnetic field $B$, thereby inducing a Zeeman shift
$\Delta E=Z\cdot M_{F}$ with an energy gap $Z=g_{F}\mu_{B}B$ between
the hyperfine states.
These states are coupled by two co-propagating Laguerre-Gauss beams with
frequencies $\omega_{1},\, \omega_{2}$ and orbital angular
momenta $\ell_{1},\, \ell_{2}$. The frequencies are chosen such that
they are detuned from any actual electronic transition, thus
single-photon transitions are suppressed.  Instead, the hyperfine
states are coupled by two-photon Raman transitions,
\cf Fig.~\ref{fig:raman}.
For every two-photon transition the atom experiences a net change in its centre of mass
angular momentum of $\Delta\ell=\ell_{1}-\ell_{2}$, while the change
in linear momentum can be neglected. Hence there exists a coupling
between $|M_{F}=-1,\ell-\Delta\ell\rangle$ and $|M_{F}=0,\ell\rangle$ as well as
between $|M_{F}=0,\ell\rangle$ and $|M_{F}=+1,\ell+\Delta\ell\rangle$.
The two light beams are slightly detuned from the Raman two-photon resonance, with
detuning  $\delta\equiv(\omega_{1}-\omega_{2})-Z/\hbar$.
Including the kinetic energies of the different angular momentum states and
 applying the rotating wave
approximation~\cite{Spielman2009}, one arrives at the full Hamiltonian, $\hat{H}(\ell)/\hbar$,
\begin{equation}
\left(
  \begin{array}{ccc}
    \frac{\hbar}{2MR^{2}}(\ell+\Delta\ell)^{2}-\delta & \Omega_{R}/2 & 0 \\
    \Omega_{R}/2 & \frac{\hbar}{2MR^{2}}\ell^{2} & \Omega_{R}/2 \\
    0 & \Omega_{R}/2 & \frac{\hbar}{2MR^{2}}(\ell-\Delta\ell)^{2}+\delta
  \end{array}
  \right).
  \label{eq:model-hamiltonian}
\end{equation}
The $\ell$ dependence has been made explicit; each atomic state is
defined by the amplitudes of the three hyperfine states and its
angular momentum $\ell$. Here $\Omega_{R}$ is the two-photon Rabi
frequency which describes the coupling between hyperfine states.  The
effect of the Zeeman splitting is given by $\delta$~\cite{footnote1}.
For each $\ell$ there are three energy eigenvalues of (\ref{eq:model-hamiltonian}),
corresponding to three  dressed energy bands.

\begin{figure}
  \centering
  \includegraphics[scale=0.7]{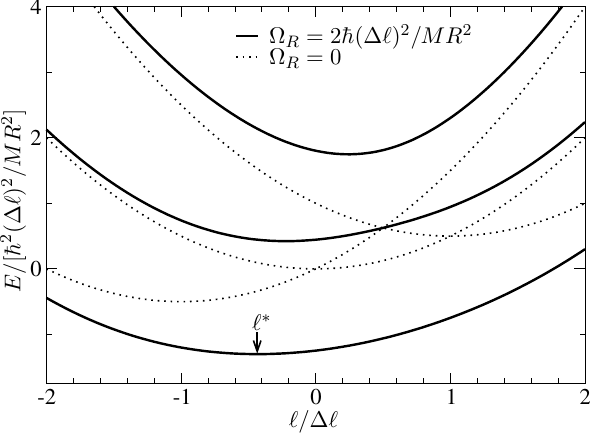}
  \caption{Coupled (solid lines) and uncoupled (dashed lines) energy bands of a three-level system, from Ref.~\protect\onlinecite{CooperHadzibabic2010}, with the parameter $\delta=0.5\hbar(\Delta\ell)^{2}/MR^{2}$. The smooth curves interpolate between the allowed integer
  values of $\ell$.}
  \label{fig:energybands}
\end{figure}

In Fig.~\ref{fig:energybands}, we show the results of a
 numerical diagonalization of the Hamiltonian~\eqref{eq:model-hamiltonian}.
The three uncoupled energy levels for $\Omega_{R}=0$ are shown as
dotted lines. When the light is on, $\Omega_R\neq 0$, these levels are
mixed and lead to the energy levels of the dressed states (solid
lines).  For non-zero detuning $\delta$ the minimum of the lowest band
is displaced to a non-zero angular momentum $\ell^{*}$.
Provided that all atoms are restricted to states in the lowest-energy
dressed band, the atoms experience an effective dispersion relation
of the form (\ref{eq:effective-dispersion}) in which the non-zero
$\ell^*$ plays the role of an azimuthal gauge field.  

Throughout this
paper, we assume that only this lowest dressed band is occupied.  This
is justified if the chemical potential $\mu$ and temperature $kT$ are
small compared to the band splitting, which is of order
$\hbar\Omega_{R}$. Hence we require that $\Omega_R$ be sufficiently
large.  
The lowest band will be referred to as $\epsilon^0_\parallel(\ell)$ in
Section~\ref{sec:interactions}. To obtain the results shown in
Section~\ref{sec:results}, we determine $\epsilon^0_\parallel(\ell)$
by numerical solution of \eqref{eq:model-hamiltonian}.
However, to allow an understanding of the general trends, we derive
some analytic expressions which are valid 
for large $\Omega_R$, where the bands are far apart from each
other and the lowest band is nearly parabolic.  Perturbation theory in
$1/\Omega_R$ shows~\cite{CooperHadzibabic2010} that the minimum of the
dispersion relation is shifted from $\ell=0$ to
\begin{equation}
  \ell^* \simeq - \sqrt{2} \frac{\delta}{\Omega_R} \Delta\ell + {\cal O}(1/\Omega_R^2) \,,
\end{equation}
and
the bare mass is increased to an effective mass $M^*$, with
\begin{equation}
  M^* \simeq M \left( 1 + \frac{\sqrt{2} \hbar(\Delta\ell)^2}{M R^2 \Omega_R} \right) + {\cal O}(1/\Omega_R^2) .
\end{equation}
Similarly,  perturbative calculations of $\left| \psi_{-1}(\ell)
\right|^2$ and $\left| \psi_{+1}(\ell) \right|^2$ show that they
have equal and opposite contributions linear in $\ell$. The
difference $\left| \psi_{-1}(\ell) \right|^2 - \left| \psi_{+1}(\ell)
\right|^2$ can indeed be written as a series in $\ell$, as in
Eq.~\eqref{eq:dp-single}, with
\begin{align}
  \Delta p_0 &\simeq \frac{\delta}{\Omega_R} \left[ \sqrt{2} - \frac{\hbar (\Delta\ell)^2}{M R^2 \Omega_R} \right] + {\cal O}(1/\Omega_R^3), \\
  \Delta p' &\simeq - \sqrt{2} \frac{\hbar\Delta\ell}{M R^2 \Omega_R} + {\cal O}(1/\Omega_R^2) .
\end{align}
As parameters representative of experiments, throughout this paper we
consider sodium with $M=23\, m_p$ (where $m_p$ is the proton mass) in
a trap of radius $R=10\,\mathrm{\mu m}$.  A typical value of the
effective mass is $M^* \approx 1.15 M$ (at a two-photon Rabi frequency
$\Omega_R=2\pi\times 100\,\mathrm{kHz}$ and for $\Delta\ell=50$).

\section{Corrections}
\label{sec:corrections}

In the preceding sections, we have summarised the theoretical proposal
of Ref.~\onlinecite{CooperHadzibabic2010}. This showed how to relate
the superfluid fraction to a spectroscopically determined hyperfine population
imbalance [see Eq.~(\ref{eq:superfluid-measurement})]. The quantitative
accuracy of this relation relies on the validity of the termination of
the Taylor expansions in Eqs.~\eqref{eq:effective-dispersion}
and~\eqref{eq:dp-single} at quadratic and linear orders respectively.
If these (terminated) expansions were exact, the superfluid
fraction would be perfectly determined by
Eq.~\eqref{eq:superfluid-measurement}. In practice, higher-order
corrections do exist, \ie
\begin{align}
  E &= E_\textrm{parabolic} + c \ell^3 + \dots \\
  \Delta p_\ell &= (\Delta p_\ell)_\textrm{linear} + c' \ell^2 + \dots
  \label{eq:dpell-corrections}
\end{align}
where $E_\textrm{parabolic}$ and $(\Delta p_\ell)_\textrm{linear}$
correspond to the lower-order expansions
\eqref{eq:effective-dispersion} and~\eqref{eq:dp-single}.

The higher-order corrections have two major implications. 

Firstly, corrections to $\Delta p_\ell$ of quadratic or higher order
in $\ell$ lead to a deviation of the spectroscopic measurement [the
right-hand side of Eq.~\eqref{eq:ell-measurement}] from the actual
average angular momentum $\langle\ell\rangle$. 

Secondly, corrections to the effective kinetic energy $E$ of cubic or
higher order in $\ell$ mean that even the very definition of the
superfluid fraction, Eq.~\eqref{eq:superfluid-definition}, breaks down.  A
basic assumption of this definition is that the rotational properties
of a (normal) gas are entirely characterised by its moment of
inertia. This is correct provided the kinetic energy is a quadratic
function of the angular momentum.  Then, under a transformation to a
frame rotating at angular frequency $\vecb{\omega}$,
the interaction energy is unchanged, and
the (parabolic) kinetic energy transforms as $E\to E' = E +
\vecb{\omega} \cdot \langle \vecb{L} \rangle + \frac{1}{2} I_\textrm{cl}
\omega^2$. Thus, for a state of given average angular momentum, the
only material property characterising the net energy change is the
moment of inertia $I_\textrm{cl}$. For non-parabolic kinetic energy,
however, the kinetic energy does not transform in any simple way, but
depends on the populations of the individual angular momentum states
and hence also on how interactions and/or temperature have distributed
particles among these levels. The moment of inertia of even a normal
gas cannot be assumed simply to be a constant $I_\textrm{cl}$.  

We now turn to estimate the quantitative effects of these two forms of
correction.

\begin{figure}[tbhp!]
  \centering
  \includegraphics[scale=0.4]{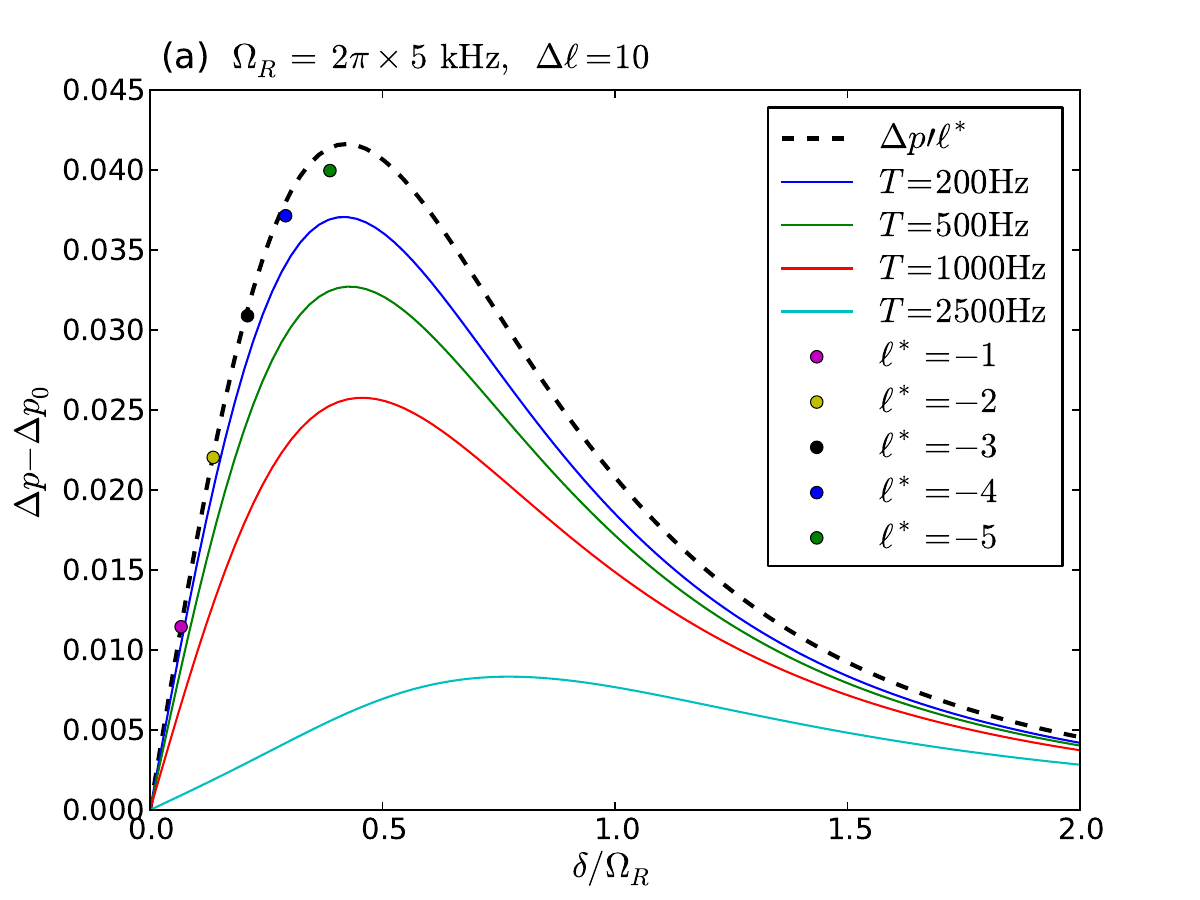}
  \includegraphics[scale=0.4]{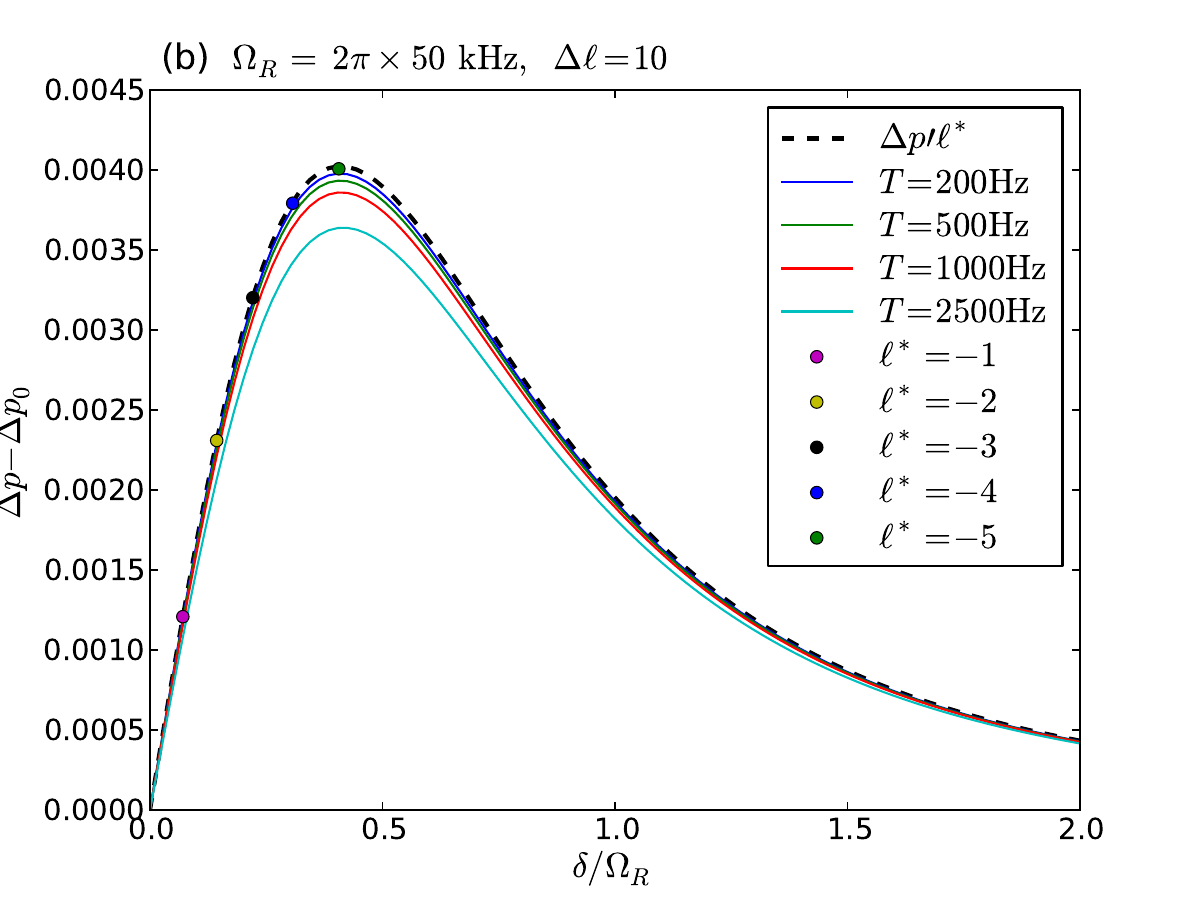}
  \includegraphics[scale=0.4]{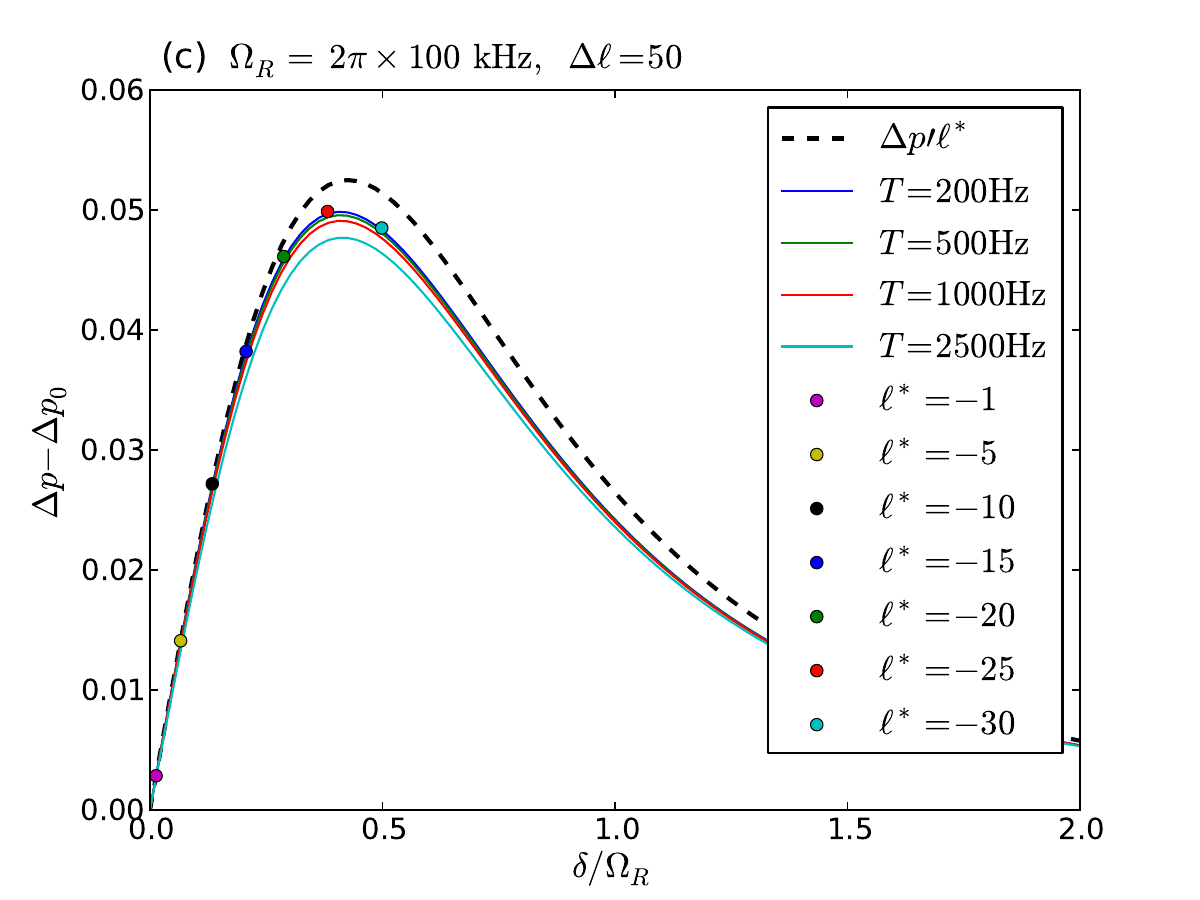}
  \caption{(Color online) Results for a normal gas: Ideal signal size $\Delta p' \ell^*$ and temperature-dependent deviation from this as a function of the detuning, $\delta/\Omega_R$. Data are for a classical Boltzmann gas at several values of $\Omega_R$ and $\Delta\ell$. Circles denote the zero-temperature limit $(\Delta p_{\ell^*} - \Delta p_0)$ at different values of $\ell^*$. The curves are labelled from top to bottom, and $\ell^*$ is labelled from left to right.}
  \label{fig:boltzmann}
\end{figure}

First we consider the effects of the higher-order corrections to
$\Delta p_\ell$, Eq.~(\ref{eq:dpell-corrections}). Because of the
departure of $\Delta p_\ell$ from its linear expansion, we find
$\Delta p' \ell^* \neq \Delta p_{\ell^*}-\Delta p_0$.
This difference is illustrated for different combinations of $\Omega_R$ and $\Delta\ell$
in Fig.~\ref{fig:boltzmann}, where the dashed lines show
$\Delta p' \ell^*$ and the circles show the actual difference
$\Delta p_{\ell^*}-\Delta p_0$. As a consequence,
Eq.~\eqref{eq:superfluid-measurement} would give a systematically
incorrect result even in the $T\to 0$ limit.
In practice this can be
corrected by replacing
the
denominator in~\eqref{eq:superfluid-measurement} by $\Delta p_{\ell^*}-\Delta p_0$.

A more serious problem arises from the fact that non-zero
temperature and/or interactions populate a range of $\ell$ states. Due to
this broadening of the distribution function, the higher-order
corrections to $\Delta p_\ell$ in Eq.~\eqref{eq:dpell-corrections}
lead to the situation that for a normal gas (or a relaxed superfluid)
with average momentum $\langle \ell\rangle = \ell^*$, the
spectroscopic signal $\Delta p$ is not just determined by the signal
of the state at $\ell^*$, but
depends on the overall distribution function, so $\Delta p \neq \Delta p_{\ell^*}$. Similarly, for the
(metastable) superfluid with  $\langle
\ell\rangle =0$, one has $\Delta p \neq \Delta p_0$.  Hence one
would incorrectly determine the superfluid fraction.
In order to estimate the size of this error, we here consider the
effect of temperature using the example of an ideal Boltzmann gas.
(Interactions will be considered in
Sections~\ref{sec:interactions} and~\ref{sec:results}.)  We populate
the different $\ell$ states in the lowest band (see Fig.~\ref{fig:energybands})
according
to a  Boltzmann distribution $n_\ell \propto
\exp[-\epsilon^0_\parallel(\ell)/kT]$. 
The results are shown for a range of temperatures by the solid lines in 
Fig.~\ref{fig:boltzmann}. If the effects of thermal broadening were
negligible, all these curves would agree with $\Delta p_{\ell^*} - \Delta p_0$
(the circles in Fig.~\ref{fig:boltzmann}). 

Increasing $\Omega_R$ decreases the relevance of higher-order
corrections,
but
also diminishes the signal size $\Delta p' \ell^*$ and thus leads to a
bigger experimental uncertainty. Increasing $\Delta\ell$ has the
opposite effect.  One goal of this paper is to find parameters
for which a good compromise can be reached, such that there is both a
large experimental signal and small systematic inaccuracies.

For $\delta/\Omega_R\ll 1$, the influence of $\Omega_R$ and
$\Delta\ell$ on signal strength and higher-order corrections can be
seen in the expressions from perturbation theory. Using the
perturbative expansions in $1/\Omega_R$
(\cf Section~\ref{sec:3-level-system}), the signal size is given by
\begin{equation}
  \Delta p' \ell^* \simeq 2 \frac{\delta}{\Omega_R} \frac{\hbar}{MR^2\Omega_R} (\Delta\ell)^2 + {\cal O}(1/\Omega_R^3) ,
\end{equation}
and considering the corrections to $\Delta p_\ell$ in~\eqref{eq:dpell-corrections}, the second-order coefficient $c'$ is given by
\begin{equation}
  c' \simeq - 3 \sqrt{2} \frac{\delta}{\Omega_R} \left( \frac{\hbar}{M R^2 \Omega_R} \right)^2 (\Delta\ell)^2 + {\cal O}(1/\Omega_R^4) .
\end{equation}
These expressions show that increasing $\Delta \ell$ is an effective
way to increase the signal size, but also has an adverse effect on the
accuracy. Accuracy is improved by increasing $\Omega_R$, but at a
reduction in signal size.
As a trade-off one would thus try to make $\Delta\ell$ as high as
possible, and then increase $\Omega_R$ as long as the signal size
remains big enough. 
As a reasonable, and experimentally feasible, compromise we choose
$\Delta\ell=50$ and $\Omega_R=2\pi\times 100\,\mathrm{kHz}$ [as in Fig.~\ref{fig:boltzmann}~(c)]. These are
the values we will use through the remainder of this paper.

Second, and finally, we return to the definition of superfluidity in
Eq.~\eqref{eq:superfluid-definition}, and estimate its
adequacy. Measuring the relative deviation $(\langle\ell\rangle -
\ell^*)/\ell^*$ is a way to assess the departure from parabolicity due
to the terms of $E$ which are odd in $\ell$. For the Boltzmann gas of
Fig.~\ref{fig:boltzmann}, the relative deviation is maximal at
$\ell^*\sim\delta\to 0$ (with a maximum value $\propto 1/\Omega_R^2$)
and decreases with higher $\delta$.  At $T=1000\,\mathrm{Hz}$~\cite{footnotet}, the
maximum value of the deviation at $\delta\to 0$ is about $0.3\%$.
To further quantify the size of
non-parabolic corrections to the dispersion relation,
we have carried out fourth-order perturbation theory~\cite{Fernandez2001}
in $1/\Omega_R$. The next terms in the series in $\ell$ are
\begin{align}
  E \simeq E_\textrm{parabolic} & - \sqrt{2} \frac{\delta}{\Omega_R} \frac{\hbar^4(\Delta\ell)^3}{(MR^2)^3 \Omega_R^2} \ell^3 \nonumber \\
  & {} + \frac{1}{2\sqrt{2}} \frac{\hbar^5(\Delta\ell)^4}{(MR^2)^4 \Omega_R^3} \ell^4 + {\cal O}(\ell^5) ,
\end{align}
where both coefficients have higher-order contributions of ${\cal
  O}(1/\Omega_R^4)$. The contributions of cubic and quartic term
relative to the parabolic term ($\hbar^2\ell^2 / 2 M^* R^2$) are of order
$10^{-4}\ell\,$ and $10^{-7}\ell^2$, respectively.  This is on the same order as
the $0.3\%$ estimated above: For $T=1000\,\mathrm{Hz}$ and $\ell^*=1$,
the root mean square angular momentum is $\sqrt{\langle\ell^2\rangle}\sim 16$, so
we would expect the relative correction due to the cubic term to be
about $0.2\%$.  This deviation of the dispersion relation from
parabolicity sets an upper bound to the accuracy with which one can
determine the superfluid fraction.

\section{Interactions}
\label{sec:interactions}

In addition to the effects of non-zero temperature on a normal fluid, we
want to know what happens to $\Delta p$ due to interactions in a
superfluid. It is important to note that a gas is only superfluid
\emph{because} of interactions; in the limit of
vanishing interaction strength, the critical velocity (see below,
Section~\ref{sec:critical-velocity}) goes to zero, so a metastable superfluid flow cannot be maintained.

\begin{figure}[tbhp]
  \centering
  \includegraphics[scale=0.4]{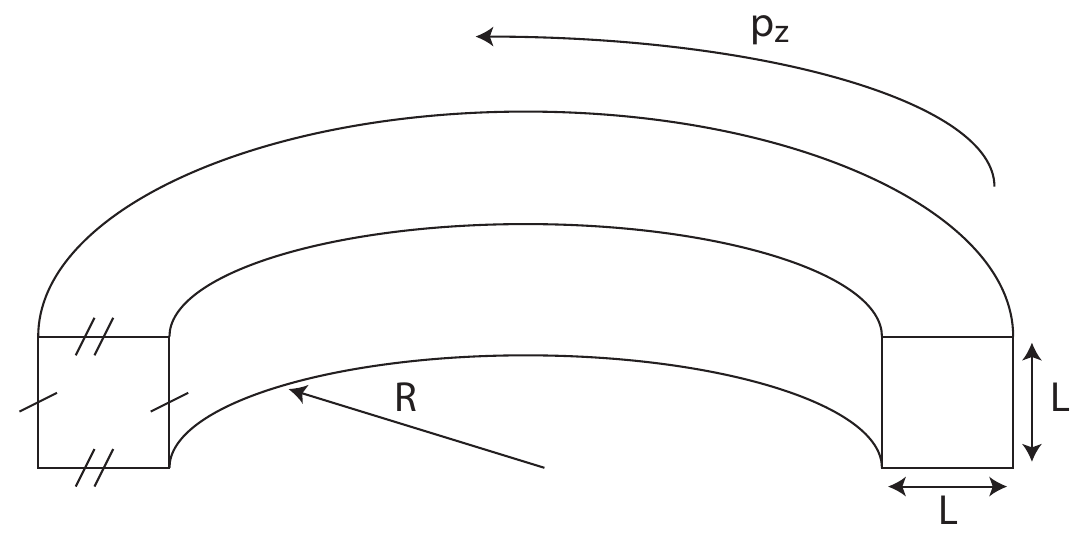}
  \caption{The system geometry employed in this section, a toroidal trap with hard walls and a cross-sectional area $L\times L$. For the radius $R \gg L$, angular momentum $\ell$ can be unrolled unto the $p_z$ direction, $p_z = \hbar\ell/R$, which then has effective periodic boundary conditions. Thus $\ell$ and $p_z$ will be used interchangeably.}
  \label{fig:sketch2}
\end{figure}

Here we do not attempt a full model of the interacting gas in a
trap. Rather, we aim to use a method that includes interactions in the
simplest way. We therefore consider an interacting gas with
\emph{uniform} density.  The density inhomogeneity near the walls is
on the scale of the healing length $\xi=(8\pi na)^{-1/2}$, which is
the distance over which the condensate recovers its bulk value from
zero density at the walls. We assume that $\xi \ll \Delta R$ and that
we can thus neglect the density inhomogeneity.
We therefore model the gas in a ring-like trap by considering a torus of radius $R$ and with a rectangular cross-sectional area $A = L \times L$, \cf Fig.~\ref{fig:sketch2}.
Since we neglect density inhomogeneities, we consider the system to be
uniform and impose periodic boundary conditions in the transverse
directions.  For $L \sim \Delta R \ll R$ we can unroll the angular
momentum from the circumferential direction of the torus onto a line
with periodic boundary conditions, equivalent to a linear momentum
$p_z = (\hbar/R) \ell$, where $\ell$ is the angular momentum (in units
of $\hbar$) defined above.  With this understanding we will use $\ell$
and $p_{z}$ interchangeably.

We are only interested in the azimuthal motion, $p_{z}$ or $\ell$, in
which we would like to measure superfluidity. Therefore we will
integrate out the two transverse directions $\vecb{p}_{\perp}$. These
have a simple kinetic energy dispersion, $p_\perp^2 / 2M$. In the
azimuthal direction we allow for a general dispersion relation
$\epsilon^0_\parallel(p_z)$, set by the energy of the lowest band of
the Hamiltonian~\eqref{eq:model-hamiltonian}.  The total dispersion
relation is given by
\begin{equation}
  \epsilon^{0}(\vecb{p}) = \frac{1}{2M} \vecb{p}_{\perp}^{2} + \epsilon_{\parallel}^{0}(p_{z}).
  \label{eq:3d-dispersion-relation}
\end{equation}
Later on we will compare the superfluid fraction with the condensate
fraction, so it is instructive to determine the number of excited
particles $N_\textrm{ex}$.  Here, we will determine this for an ideal
non-interacting gas, as many of the features will carry over to the
interacting case discussed below.  In a non-interacting gas the number
of excited particles is given by
\begin{equation}
  N_{\text{ex}}=\sum_{\vecb{p}\neq0}n_B\left(\epsilon^{0}(\vecb{p})\right),
\end{equation}
where $n_B(\epsilon) = [\exp(\epsilon/kT) - 1]^{-1}$ is the Bose-Einstein distribution in the condensed system (for which the chemical potential vanishes, $\mu = 0$).
We split the sum,
\begin{equation}
  N_{\text{ex}} = N_{\text{ex}}^{\perp}(p_z=0) + \sum_{p_{z}\neq0}N_{\text{ex}}^{\perp}(p_{z}) ,
\end{equation}
defining
\begin{equation}
  N_{\text{ex}}^{\perp}(p_{z}) \equiv \sum_{\vecb{p}_{\perp}\neq0} n_B\left(\vecb{p}_{\perp}^{2}/2M + \epsilon_{\parallel}^{0}(p_{z})\right) .
\end{equation}
Here we assume without loss of generality that the energy minimum is at $\epsilon_{\parallel}^{0}(p_{z}=0) = 0$.
We switch to an integral representation according to $\sum_{p} \to \frac{L}{2\pi\hbar} \int dp$. The term $N_\textrm{ex}^\perp (p_z)$ with $p_z\neq 0$ does not pose any problems. However, when evaluating the $p_{z}=0$ contribution, we find
\begin{align}
  N_{\text{ex}}^{\perp}(p_{z}=0) &= \frac{A}{(2\pi\hbar)^{2}} \int d^{2}\vecb{p}_{\perp} n_B(\vecb{p}_{\perp}^{2}/2M) \nonumber \\
  &= \frac{2\pi A}{(2\pi\hbar)^{2}} \int_{0}^{\infty} dp\, p\, n_B(p^{2}/2M),
  \label{eq:nex-pz0-term}
\end{align}
which is infrared divergent. For $p\to 0$, the integrand is
approximately given by $p/(p^{2}/2M\,kT) \sim 1/p$, thus the
indefinite integral at small momenta is $\sim-\log(p_{\rm min})$. Fortunately, in
the thermodynamic limit this logarithmic divergence is unproblematic.
Noting that the 
low-momentum cut-off is
$p_{\text{min}}\sim\frac{1}{L}$,
one finds $N_\text{ex}^{\perp}(p_{z}=0) \sim (A/\lambda_T^2)\log(L/\lambda_T)$,
where $\lambda_T$ is the thermal de Broglie wavelength.
The other contributions are extensive,
$N_\text{ex}^{\perp}(p_{z}\neq 0)  \sim (V/\lambda_T^3)$, so
\begin{equation}
  \frac{N_\text{ex}^{\perp}(p_{z}=0)}{N_\text{ex}^{\perp}(p_{z}\neq 0)} \sim \frac{\lambda_T}{R} \log(L/\lambda_T) ,
\end{equation}
and $N_\text{ex}^{\perp}(p_{z}=0)$ can be neglected in the limit $\lambda_T \ll R$.
The same reasoning applies when taking interactions into account, so from now on we will drop the $p_{z}=0$ contribution to $N_{\text{ex}}$ and replace $\sum_{\vecb{p} \neq 0}$ by $\sum_{p_{z} \neq 0} \sum_{\vecb{p}_{\perp} \neq 0}$.

\subsection{Weak interactions with asymmetric dispersion}

In the following we consider an interacting gas, with a general,
possibly asymmetric, non-interacting dispersion relation
$\epsilon^{0}(\vecb{p})$.  Since we study atoms in the ultracold
limit, all interactions can be treated as contact interactions,
$U_{\text{eff}}(\vecb{r}, \vecb{r'}) = U_{0}\delta(\vecb{r} -
\vecb{r'})$. Here we take $U_{0}=4\pi\hbar^{2}a/M$, assuming for
simplicity that the scattering length $a$ is independent of the
internal state. The Hamiltonian in the momentum basis is
\begin{align}
  \hat{H} = & \sum_{i,\vecb{p}} \epsilon_{i,\vecb{p}}^{0} \hat{a}_{i,\vecb{p}}^{+} \hat{a}_{i,\vecb{p}} \nonumber \\
  & {} + \frac{1}{2V} U_0 \sum_{ijkl} \sum_{\vecb{p},\vecb{p}',\vecb{q}} \hat{a}_{i,\vecb{p}+\vecb{q}}^{+} \hat{a}_{j,\vecb{p}'-\vecb{q}}^{+} \hat{a}_{k,\vecb{p}'} \hat{a}_{l,\vecb{p}} .
  \label{eq:general-weakly-interacting-hamiltonian}
\end{align}
The indices $i,j,k,l$ stand for internal states, \eg one of the three hyperfine bands.
As we assume that only the lowest band is occupied (\cf Section~\ref{sec:3-level-system}), we can ignore inter-band mixing and simplify Eq.~\eqref{eq:general-weakly-interacting-hamiltonian} to the Hamiltonian we will consider hereafter:
\begin{equation}
  \hat{H} = \sum_{\vecb{p}} \epsilon^{0}(\vecb{p}) \hat{a}_{\vecb{p}}^{+} \hat{a}_{\vecb{p}} + \frac{U_{0}}{2V} \sum_{\vecb{p},\vecb{p}',\vecb{q}} \hat{a}_{\vecb{p}+\vecb{q}}^{+} \hat{a}_{\vecb{p}'-\vecb{q}}^{+} \hat{a}_{\vecb{p}'} \hat{a}_{\vecb{p}} .
  \label{eq:approximate-contact-hamiltonian}
\end{equation}

\subsubsection{Bogoliubov transformation}

We assume that the condensate has only one macroscopically occupied
state $\vecb{p}_c=0$.  Should the condensate be in a state $\vecb{p}_c
\neq 0$, we can shift all momenta by relabelling the states to
$\vecb{p}' \equiv \vecb{p} - \vecb{p}_c$ with energy
$\epsilon^{0\prime}(\vecb{p}') = \epsilon^0(\vecb{p}_c+\vecb{p}')$.  We
note that, for a superfluid, the state $\vecb{p}_c$ is not necessarily
the ground state: for example, as in the protocol described above, a
superfluid may remain condensed in $\ell=0$ even after a gauge field
is applied such that the lowest-energy single particle state has
$\ell^*\neq 0$. Our discussion in this section also applies to
these metastable superfluid states.

The creation and annihilation operators of the ground state are
$\hat{a}_0^+$ and $\hat{a}_0$, respectively. We have $\langle
\hat{a}_0^+ \hat{a}_0 \rangle = N_0$, where $N_0$ is the number of
particles in the condensate. For $N_0 \gg 1$, we can therefore neglect
the commutator $[ \hat{a}_{0}, \hat{a}_{0}^{+} ]_{-} = 1$
compared to the operator $\hat{a}_{0}^{+} \hat{a}_{0}$, and replace
the ground state operators by a c-number: $\hat{a}_{0}\simeq
\hat{a}_{0}^{+}\simeq\sqrt{N_{0}}$. Expanding the interaction term and
only keeping terms of order $\mathcal{O}(N_{0})$ or higher (\ie
assuming the only relevant interactions are with the condensate
state), one finds
\begin{eqnarray}
  \hat{H} & \simeq & N \epsilon^{0}(0) + \frac{U_{0}N^{2}}{2V} \nonumber \\
  &  & {} + \sum_{\vecb{p}\ne0} \left[\epsilon^{0}(\vecb{p}) - \epsilon^{0}(0) + U_{0}n_{0}\right] \hat{a}_{\vecb{p}}^{+} \hat{a}_{\vecb{p}} \nonumber \\
  &  & {} + \frac{U_{0}n_{0}}{2} \sum_{\vecb{p}\ne0} \left(\hat{a}_{\vecb{p}}^{+} \hat{a}_{-\vecb{p}}^{+} + \hat{a}_{\vecb{p}} \hat{a}_{-\vecb{p}}\right) \nonumber \\
  & = & \sum_{\vecb{p}\ne0}' \left\{ [\epsilon^{0}(\vecb{p}) - \epsilon^{0}(0) + \epsilon_{1}] \, \hat{a}_{\vecb{p}}^{+} \hat{a}_{\vecb{p}} \right. \nonumber \\
  &   & \left. {} + [\epsilon^{0}(-\vecb{p}) - \epsilon^{0}(0) + \epsilon_{1}] \, \hat{a}_{-\vecb{p}}^{+} \hat{a}_{-\vecb{p}} \right\} \nonumber \\
  &  & {} + \sum_{\vecb{p}\ne0}' \epsilon_{1} \left(\hat{a}_{\vecb{p}}^{+} \hat{a}_{-\vecb{p}}^{+} + \hat{a}_{\vecb{p}} \hat{a}_{-\vecb{p}}\right) + E_{\text{offset}} ,
  \label{eq:bogoliubov-hamiltonian}
\end{eqnarray}
where the prime on the sum denotes that we only sum over (an arbitrary) half of momentum space. Here $n_0 \equiv N_0/V$ is the condensate density, and we have defined $\epsilon_{1} \equiv U_{0}n_{0} = 4\pi\hbar^2 a n_0/M$, which sets the chemical potential, $\mu=\epsilon_1$. The constant terms have been absorbed into $E_{\text{offset}}$.

We diagonalize the Hamiltonian using the Bogoliubov transformation
\begin{align}
  \hat{\alpha}_{\vecb{p}} &:= u \hat{a}_{\vecb{p}} + v \hat{a}_{-\vecb{p}}^{+}, \nonumber \\
  \hat{\alpha}_{-\vecb{p}} \equiv \hat{\beta}_{\vecb{p}} &:= u \hat{a}_{-\vecb{p}} + v \hat{a}_{\vecb{p}}^{+} ,
\end{align}
with $[\hat{\alpha},\hat{\alpha}^{+}]_{-} = [\hat{\beta},\hat{\beta}^{+}]_{-} = u^{2}-v^{2} = 1$.
The Hamiltonian~\eqref{eq:bogoliubov-hamiltonian} is diagonalized by choosing
$u^2 = \frac{1}{2} \left( \frac{\bar{\epsilon}}{\epsilon} + 1 \right)$ and
$v^2 = \frac{1}{2} \left( \frac{\bar{\epsilon}}{\epsilon} - 1 \right)$,
leading to
\begin{equation}
  \hat{H} = E_{\text{offset}} + \sum_{\vecb{p}\neq0} \frac{1}{2} \left(\epsilon-\bar{\epsilon}\right) + \sum_{\vecb{p} \neq 0} \underbrace{\left(\epsilon+\bar\gamma\right)}_{\epsilon_{\text{exc}}} \hat{\alpha}_{\vecb{p}}^{+} \hat{\alpha}_{\vecb{p}} ,
\end{equation}
where $\epsilon_\text{exc}(\vecb{p}) = \epsilon(\vecb{p})+\bar\gamma(\vecb{p})$ is the energy of Bogoliubov excitations. We 
have defined~\cite{footnote3}
\begin{align}
  \gamma(\vecb{p}) & = \frac{1}{2}\left[\epsilon^{0}(\vecb{p})+\epsilon^{0}(-\vecb{p})\right] - \epsilon^{0}(0),\nonumber \\
  \bar\gamma(\vecb{p}) & = \frac{1}{2}\left[\epsilon^{0}(\vecb{p})-\epsilon^{0}(-\vecb{p})\right],\nonumber \\
  \bar{\epsilon}(\vecb{p}) & = \gamma(\vecb{p})+\epsilon_{1},\nonumber \\
  \epsilon(\vecb{p}) & = \sqrt{\bar{\epsilon}^{2}-\epsilon_{1}^{2}} = \sqrt{\gamma(\gamma+2\epsilon_{1})}.
  \label{eq:bogoliubov-parameters}
\end{align}
$\gamma(\vecb{p})$ denotes the symmetric part of the non-interacting dispersion relation, shifted such that $\gamma(\vecb{p}_c)=0$. $\epsilon(\vecb{p})$ is the symmetric part of the Bogoliubov excitation energy.
$\bar\gamma(\vecb{p})$ denotes the asymmetric part of the excitation energy, and is the same
for both non-interacting and interacting cases.
As opposed to the standard case of a symmetric dispersion relation, we
find $\bar\gamma \neq 0$ and a dependence of the energy
$\epsilon_\textrm{exc}$ of Bogoliubov excitations on $\ell^*$. We
shall use this in Section~\ref{sec:critical-velocity} to determine
the critical velocity at which the metastable superfluid becomes
unstable.

\subsubsection{Depletion of the condensate}

We can write the operator for the total number of particles as
\begin{equation}
  \hat{N} = N_{0}+\sum_{\vecb{p}\ne0} \hat{a}_{\vecb{p}}^{+} \hat{a}_{\vecb{p}}.
\end{equation}
When we apply the Bogoliubov transformation, we find
\begin{equation}
  N \simeq N_{0} + \sum_{\vecb{p}\ne0} \frac{1}{2}\left(\frac{\bar{\epsilon}}{\epsilon}-1\right) + \sum_{\vecb{p}\ne0} \frac{\bar{\epsilon}}{\epsilon}\left\langle \hat{\alpha}_{\vecb{p}}^{+}\hat{\alpha}_{\vecb{p}}\right\rangle ,
\end{equation}
where we have already taken the expectation value.
At zero temperature,
\begin{equation}
  N(T=0) = N_{0} + \sum_{\vecb{p}\ne0} \frac{1}{2}\left( \frac{\bar{\epsilon}}{\epsilon} - 1 \right) .
\end{equation}
After changing the sum to an integral, we can analytically integrate out the transverse directions. In terms of the number density $n = N/V$ we get
\begin{equation}
  n(T=0) = n_{0} + \frac{n_{0}a}{2\pi R} \sum_{p_{z}\neq0} \left[ 1 + \frac{\gamma}{\epsilon_{1}} \left( 1 - \sqrt{1+2\frac{\epsilon_{1}}{\gamma}} \right)\right],
  \label{eq:int-dist-zero}
\end{equation}
where $\gamma$ is evaluated at $\vecb{p}_{\perp}=0$; $\gamma = \gamma(p_z; \vecb{p}_\perp=0)$. Each term of the sum gives the corresponding $\ell$-state occupation $n_\ell (T=0)$.

To find the temperature-dependent depletion, we calculate the expected excitation number: $\langle \hat{\alpha}_{\vecb{p}}^{+} \hat{\alpha}_{\vecb{p}} \rangle = n_B\left(\epsilon_{\text{exc}}(\vecb{p})\right)$. The energy of such thermal excitations is given by $\epsilon_{\textrm{exc}} = \epsilon + \bar\gamma$. Changing the integral over transverse directions into an integration over the dimensionless $x = p_\perp^2 / (2 M \epsilon_1)$ leads to
\begin{align}
  & n(T)-n(T=0) \nonumber \\
  & = \frac{n_{0}a}{\pi R} \sum_{p_{z}\neq0} \int_{0}^{\infty} dx \frac{x+1+\gamma/\epsilon_{1}}{\sqrt{(x+1+\gamma/\epsilon_{1})^{2}-1}} \times \nonumber \\
  & \quad \times \frac{1}{\exp\left[\left(\epsilon_{1}\sqrt{(x+1+\gamma/\epsilon_{1})^{2}-1}+\bar\gamma\right)/kT\right]-1} ,
  \label{eq:int-dist-temp}
\end{align}
again with $\gamma = \gamma(p_{z}; \vecb{p}_{\perp}=0)$.
In general, this integral has to be evaluated numerically. Similarly to above, the terms of this sum give the finite-temperature distribution function $n_\ell (T) - n_\ell (T=0)$.

\subsection{Popov approximation}

To extend the validity of our approximation to higher temperatures, we
employ the Popov approximation.
In the homogeneous system, the Popov approximation leads to the same Hamiltonian as in the Bogoliubov theory, except for a simple excitation-independent energy offset. Thus the distribution function stays the same --- the only difference is that now we need to determine $n_0$ self-consistently, both at zero and at non-zero temperature~\cite{PethickSmith2008,Popov1988Functional}.

As the zero-temperature distribution~\eqref{eq:int-dist-zero} is symmetric around $\ell=0$, at $T=0$ the average angular momentum is always zero, $\langle\ell\rangle \equiv 0$, for any interaction strength.
This recovers the expectation that $\rho_s/\rho=1$ even if the condensate is depleted by interactions.
We calculate the zero-temperature depletion for a parabolic dispersion relation $\epsilon^0_\parallel = \frac{\hbar^2\ell^2}{2MR^2} = \gamma$. We approximate the sum in Eq.~\eqref{eq:int-dist-zero} by an integral,
\begin{equation}
  \frac{n-n_0}{n_0} = \frac{a}{2\pi R}\frac{2\pi R}{2\pi\hbar} \int_0^\infty dp_z \left[ 1 + \frac{\gamma}{\epsilon_{1}} \left( 1 - \sqrt{1+2\frac{\epsilon_{1}}{\gamma}} \right)\right] ,
\end{equation}
which can be evaluated analytically, giving
\begin{align}
  \frac{n-n_0}{n_0} &= \frac{a\sqrt{2M}}{2\pi\hbar} \sqrt{\frac{4\pi \hbar^2 a n_0}{M}} \left(+\frac{4}{3}\sqrt{2}\right) \nonumber \\
  &= \frac{8}{3\sqrt{\pi}} \left(n_0 a^3\right)^{1/2} ,
\end{align}
where we substituted $\epsilon_1=4\pi\hbar^2 a n_0 / M$. The resulting expression for the condensate fraction,
\begin{equation}
  \label{eq:n0-popov}
  \frac{n_0}{n} \simeq \left[ 1 + \frac{8}{3\sqrt{\pi}} \left(n_0 a^3\right)^{1/2} \right]^{-1} ,
\end{equation}
is not in a closed form; we have to find a self-consistent solution for $n_0$ numerically. At small depletions $n_\textrm{ex}/n \ll 1$, Eq.~\eqref{eq:n0-popov} simplifies to the Bogoliubov result~\cite{PethickSmith2008}, $n_0/n \simeq 1 - \frac{8}{3\sqrt{\pi}} (na^3)^{1/2}$.

\vspace{2ex}

According to this theory, one would simply occupy the lowest band as
stated in the interacting distribution function given by
Eq.~\eqref{eq:int-dist-temp} until convergence is achieved.  For a
parabolic dispersion relation, Eq.~\eqref{eq:int-dist-temp} has an
asymptotic behaviour $n_\ell \propto 1/\ell^2$ at large momenta, $\ell
\gg 1$, and thus extends to very high $\ell$.  However, the
\emph{physical} distribution of atoms does not extend beyond
$\ell_\textrm{max} \sim R/R_e$, where $R_e$ is the range of the
interaction potential (in practical terms, this is of the same order
as the scattering length $a$).  Thus we do not trust the expression
for $n_\ell$, Eq.~\eqref{eq:int-dist-temp}, for $\ell/R \gtrsim 1/R_e
\sim 1/a$, and so it is pointless to do the whole sum. In any case,
all the important physics characterising the superfluid response is
contained in $p \lesssim \hbar/\xi$ (equivalent to $\ell \lesssim
R/\xi$). Hence we introduce a cut-off by ignoring states with $\ell >
\ell_{\rm cut}$. We choose $\ell_{\rm cut} \sim 10\times 2\pi R/\xi$,
for which the relative error in $\dpRSF - \dpSF$ due to the cut-off
with respect to $\ell_\textrm{cut} = \infty$ is on the order of
$10^{-4}$.

\subsection{Critical velocity}
\label{sec:critical-velocity}

In order that the system is (meta)stable, the minimum of the
excitation energy $\epsilon_{\text{exc}} = \epsilon+\bar\gamma$ has to be
positive: $\min_\ell \epsilon_{\text{exc}}(\ell) > 0$. This leads to a
critical angular momentum shift $\ell^*_\textrm{crit}$ above which, for
$|\ell^*| > \ell^*_\textrm{crit}$, the superfluid flow is
unstable. (This is the Landau criterion for stable superfluid flow.)
The determination of the critical $\ell^*$ in the general case
requires a full (numerical) determination of the dispersion relation
of the lowest dressed band, $\epsilon^0_\parallel(p_z)$. In the limit
in which the non-interacting dispersion relation can be taken to be
parabolic [\cf Eq.~\eqref{eq:effective-dispersion}], the symmetric and
antisymmetric parts of $\epsilon^0_\parallel$ are given by $\gamma
(\ell) = \frac{\hbar^2}{2 M^* R^2} \ell^2$ and $\bar\gamma (\ell) = -
\frac{\hbar^2 \ell^*}{M^* R^2} \ell$.
Then the lowest-energy Bogoliubov excitation at given $\ell$ is
\begin{equation}
  \epsilon_{\rm exc}(\vecb{p}_\perp = 0) = \epsilon + \bar\gamma \simeq \sqrt{\frac{\hbar^2 \epsilon_1}{M^* R^2}} |\ell| - \frac{\hbar^2 \ell^*}{M^* R^2} \ell\,.
\end{equation}
This is positive for all $\ell$ only if $$\hbar^2 |\ell^*| / M^* R^2 < \sqrt{\hbar^2 \epsilon_1 / M^* R^2}\,.$$ Substituting $\epsilon_1 = 4\pi\hbar^2 a n_0 / M$, this corresponds to
\begin{equation}
  |\ell^*| < \ell^*_{\rm crit} = \sqrt{4\pi R^2 a n_0 \frac{M^*}{M}} \, .
\label{eq:lcrit}
\end{equation}
If $|\ell^*| < \ell^*_{\rm crit}$, then $\epsilon_{\rm exc}(\ell) > 0$
for all $\ell$. On the other hand, if $|\ell^*| > \ell^*_{\rm crit}$,
then there always is some $\ell$ such that $\epsilon_{\rm exc}(\ell) <
0$ and hence the system is thermodynamically unstable. It will relax due to the
introduction of vortices.  Eq.~(\ref{eq:lcrit}) shows that the
metastable superfluid will become unstable when the condensate density
is too low (\ie when the temperature is too high), when interactions
are too weak, or when $\ell^*$ is too large.

\section{Results}
\label{sec:results}

In order to present the results of our calculations for an interacting
gas, we consider two scenarios. In the first one, we study a Bose-Einstein condensate (BEC) with
weak interactions, Section~\ref{subsec:qualitative}, for which we show that
the proposed measurement technique gives the expected feature that
the superfluid density and condensate density are almost the same. In
the second scenario, Section~\ref{subsec:quantitative}, we turn to the more
interesting case of a BEC with strong interactions, for which the
condensate fraction and superfluid fraction are significantly
different from each other.  We show that the measurement protocol can
distinguish the superfluid fraction from the condensate
fraction. Finally, we will analyse the trade-off between signal
strength and accuracy in Section~\ref{subsec:tradeoff}.

As described above, in our studies the condensate is assumed to be
uniform and boundary effects are neglected. When discussing
temperature dependences, we shall also assume that the density $n$ is
kept constant. For a harmonically trapped gas, with falling temperature the peak density
quickly rises by more than an order of magnitude from near the
transition point to when most of the atoms are in the condensate. We
ignore this effect, since we are primarily interested in the low
temperature properties of the BEC and less in the behaviour around
the transition point.  In the results we present, we choose the following
parameters: $M = 23 \, m_p$, $R = 10 \, \mathrm{\mu m}$, $\Omega_R =
2\pi\times 100 \, \mathrm{kHz}$, $\Delta\ell = 50$, and $n = 10^{14}
\, \mathrm{cm^{-3}}$. These are typical parameters achievable with
current technology.

\subsection{Weakly interacting BEC}
\label{subsec:qualitative}

For a new experimental technique, it is important to establish that it
works in a situation where the result is known. To that end, it will
be valuable to see the results for a weakly interacting BEC, which can
be well described by mean-field theories. Here we present the signal
that one can expect to measure in such an experiment, for a weakly
interacting BEC with $na^3=10^{-4}$. The condensate fraction at
  $T=0$ is approximately $99 \%$, showing negligible depletion.

As described in Section~\ref{sec:set-up}, the appearance of superfluidity
displays a clear {\it qualitative} signal in this measurement technique.
The processes of cooling and of imposing non-zero gauge field $\ell^*$
do not ``commute'' for a superfluid, so its response is
hysteretic. First introducing $\ell^*$ (by increasing the Raman detuning $\delta$) and
then cooling the gas leads to a condensate in the new ground state,
$\ell_c=\ell^*$. This is the ``relaxed superfluid'' (RSF), with
spectroscopic signal $\dpRSF$.  On the other hand, cooling to below
the transition point at $\ell^*=\delta=0$ creates the superfluid in
$\ell_c=0$. Subsequently increasing $\delta$ leads to a metastable
(non-relaxed) superfluid (SF), with spectroscopic signal $\dpSF$.
Hence it is possible to measure two separate curves, $\dpRSF$
respectively $\dpSF$.  In contrast, for a normal fluid, cooling and
increasing $\delta$ do commute.

\begin{figure}[tbhp]
  \centering
  \includegraphics[scale=0.4]{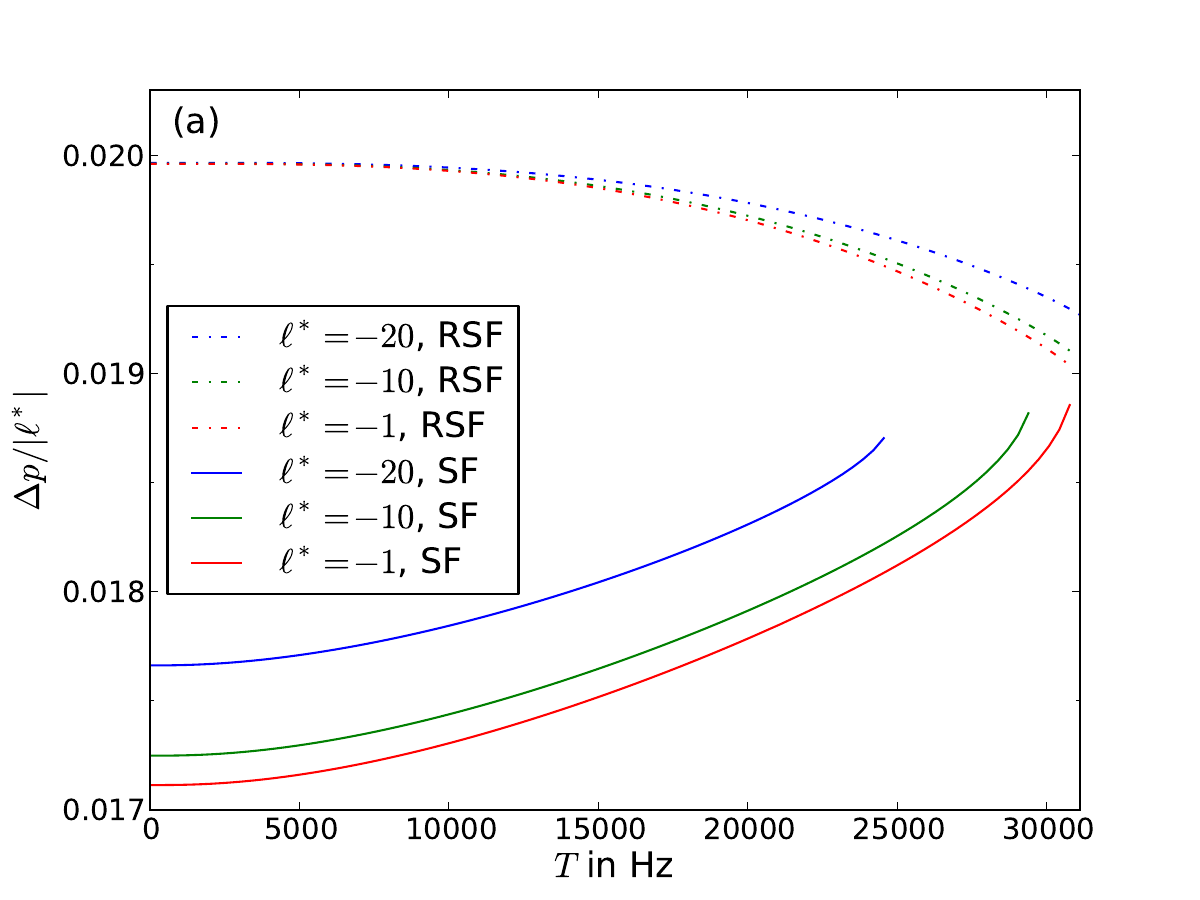}
  \includegraphics[scale=0.4]{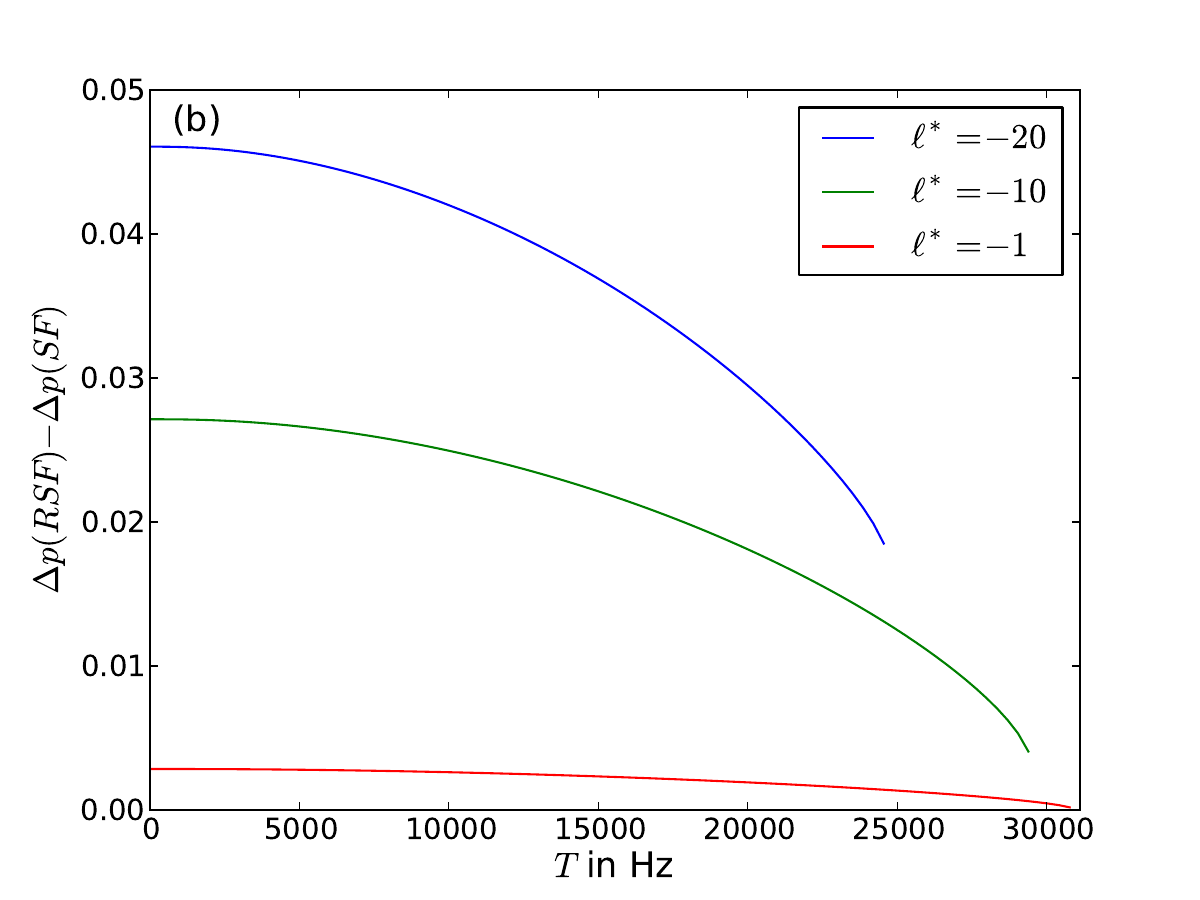}
  \includegraphics[scale=0.4]{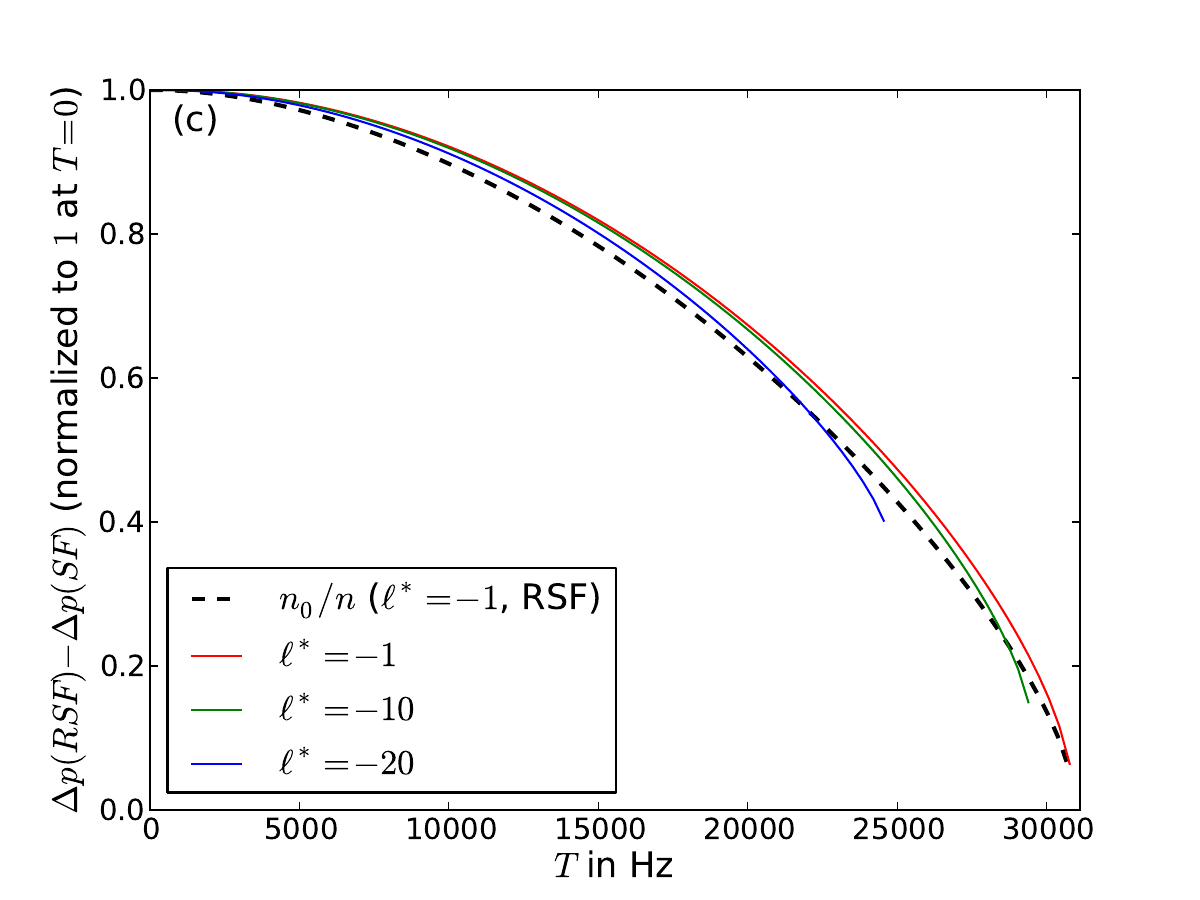}
  \caption{(Color online) Qualitative behaviour of the proposed superfluidity
    measurement as a function of temperature (in terms of the \emph{ideal} gas transition temperature $T_c\approx 31.1\,\mathrm{kHz}$) at $na^3=10^{-4}$:
    (a) the raw $\Delta p$ signal scaled by $|\ell^*|$;
    (b) difference signal $\dpRSF-\dpSF$;
    (c) difference signal normalised to
    $1$ at $T=0$, and comparison with the condensate fraction
    $n_0/n$. Curves break off due to reaching $|\ell^*| >
    \ell^*_\textrm{crit} \simeq \sqrt{4\pi R^2 a n_0}$. The curves are labelled from top
    to bottom.}
  \label{fig:hysteresis}
\end{figure}

In Fig.~\ref{fig:hysteresis}~(a), we show the spectroscopic signatures of the metastable
superfluid ($\dpSF$) and the relaxed superfluid ($\dpRSF$) at
temperatures below the BEC transition temperature $T_c$.  (The raw signal
has been scaled by $\ell^*$ so that the numbers are of the same
order.)  One can see that the curves of the metastable superfluid
($\dpSF$) terminate at certain temperatures.
 This termination happens at a temperature (below the transition
temperature) when the
condensate becomes sufficiently depleted so that $|\ell^*| >
\ell^*_\textrm{crit}$. Thus curves with higher $|\ell^*|$
break off at lower temperatures.
The difference in the signals $\dpRSF\neq \dpSF$ below
this temperature is a clear signature of the metastable superfluid
flow.

In order to extract a quantitative measure of the superfluid fraction
from these results, one could take the curves for the metastable
superfluid $\dpSF$ and use these in
Eq.~(\ref{eq:superfluid-measurement}). Application of Eq.~(\ref{eq:superfluid-measurement})
requires  knowledge of $\Delta p_0$ and $\Delta p'$.  As we have
discussed, the result will also include some quantitative corrections from higher-order terms in the Taylor expansions of $E_\ell$ and $\Delta p_\ell$
(\ref{eq:dpell-corrections}). In view of these facts, it is helpful to
take the difference $\dpRSF - \dpSF$. This removes the dependence on
$\Delta p_0$ and also removes additive systematic errors. 
The differences $\dpRSF - \dpSF$ are shown in
Fig.~\ref{fig:hysteresis}~(b).
Finally, we scale these differences by the value of the same quantity
that is obtained at $T=0$ for a weakly interacting gas,
$$\dpRSF(T=0,na^3\approx 0) - \dpSF(T=0,na^3\approx 0) \,.$$
[Since here we consider $na^3 = 10^{-4}$,  this is almost identical
to scaling by the $T=0$ limit of the curves in Fig.~\ref{fig:hysteresis}~(b).]
These measurements are for states in which we know the system should
be completely condensed and perfectly superfluid at $T=0$. They provide a
direct measurement of the quantity $\Delta p'\ell^*$, while also
removing some multiplicative systematic errors.  

The final scaled curves are shown in
Fig.~\ref{fig:hysteresis}~(c). These show what the spectroscopic
measurement would give for the superfluid fraction of the weakly
interacting BEC as a function of temperature. The different values of
$\ell^*$ correspond to different effective rotation rates.
For comparison, the dashed line shows the numerically determined {\it
  condensate fraction} $n_0/n$~\cite{footnote2}. In this case, of a
weakly interacting BEC, it is expected that the condensate and
superfluid fractions should coincide.  Fig.~\ref{fig:hysteresis}~(c)
shows that this result is recovered to a very good accuracy by the
spectroscopic measurement of the superfluid fraction.  Furthermore, in
a practical experimental measurement with a harmonically trapped gas we expect to find even better
agreement between the spectroscopic measurement and $\rho_s/\rho$:
here we chose $n=10^{14}\,\mathrm{cm^{-3}}$, which is a representative
value for the density in the BEC, but this leads to a transition
temperature which is several times higher than that found in
experiments. In a typical experiment, the density would vary such that
$T_c$ is lower, so $kT/\hbar\Omega_R$ would be smaller and
quantitative corrections should have less of an effect.

\subsection{Strongly interacting BEC}
\label{subsec:quantitative}

We will now consider how the spectroscopic method can be used to
provide an accurate quantitative measurement of the superfluid
fraction in a regime where the condensate fraction $n_0/n$ and the
superfluid fraction are significantly different. To this end, we shall
consider a relatively strongly interacting BEC, with an interaction
parameter up to $na^3=0.1$.  We cannot trust Popov theory to be
an accurate quantitative theory of an atomic gas in regimes of strong
interactions. However, we can still use the results of Popov theory to
assess the accuracy of the spectroscopic method of measuring the
superfluid fraction.  Specifically, we know that, even for a strongly
interacting gas, in the low temperature limit the superfluid fraction
should be $\rho_s/\rho = 1$, while the condensate fraction can be
significantly depleted. Here we wish to demonstrate that the
difference between condensate fraction and superfluid fraction can be
observed using the spectroscopic measurement method.

As described in Section~\ref{sec:corrections}, while considering an
accurate quantitative measurement of $\rho_s/\rho$, we face problems
coming from higher-order corrections, which even cause the usual
definition of superfluidity to break down. For this reason, in order
to proceed we follow the protocol that was outlined in
Section~\ref{subsec:qualitative}. We propose to {\it define} a
spectroscopically measured superfluid fraction by 
\begin{equation}
  \frac{\rho_s^\text{spec}}{\rho} \equiv \frac{\dpRSF(T,na^3) - \dpSF(T,na^3)}{\dpRSF(T=0,na^3\approx 0) - \dpSF(T=0,na^3\approx0)} .
  \label{eq:new-sf-measurement}
\end{equation}
The denominator gives the reference signal at zero temperature and
with negligible interactions, where the system is completely condensed
into the ground state and $100\%$ superfluid.  (The interactions
cannot be exactly zero, since one requires $|\ell^*| <
\ell^*_\textrm{crit} \simeq \sqrt{4\pi R^2 a n_0}$ for metastability
of the superfluid flow.)  In principle, all four $\Delta p$ values in
Eq.~(\ref{eq:new-sf-measurement}) can be measured in one experimental
system by using a Feshbach resonance~\cite{Chin2010} to tune $a$. All measurements
need to be taken at the same values for $\Omega_R$, $\Delta\ell$ and
$\ell^*$. One should take the limit $\ell^*\to 0$, analogous to
$\omega\to 0$ in the definition~\eqref{eq:superfluid-definition}.

As described in Section~\ref{sec:corrections}, all higher-order corrections vanish for $\Omega_R \to \infty$, so in this limit the
definition~\eqref{eq:superfluid-definition} coincides 
with the spectroscopic definition~\eqref{eq:new-sf-measurement}. Thus,
the superfluid fraction is obtained from the spectroscopic signal
(\ref{eq:new-sf-measurement}) by taking the limit
\begin{equation}
  \frac{\rho_s}{\rho} \equiv \lim_{\Omega_R\to\infty} \left(\frac{\rho_s^\text{spec}}{\rho}\right) .
\end{equation}
Evaluating Eq.~\eqref{eq:new-sf-measurement} at different values of
$\Omega_R$ gives a way to assess the influence of non-parabolic
corrections. Furthermore, this offers the possibility of improving the
measurement by extrapolating to $\Omega_R\to\infty$.

\begin{figure}[tbhp]
  \centering
  \includegraphics[scale=0.4]{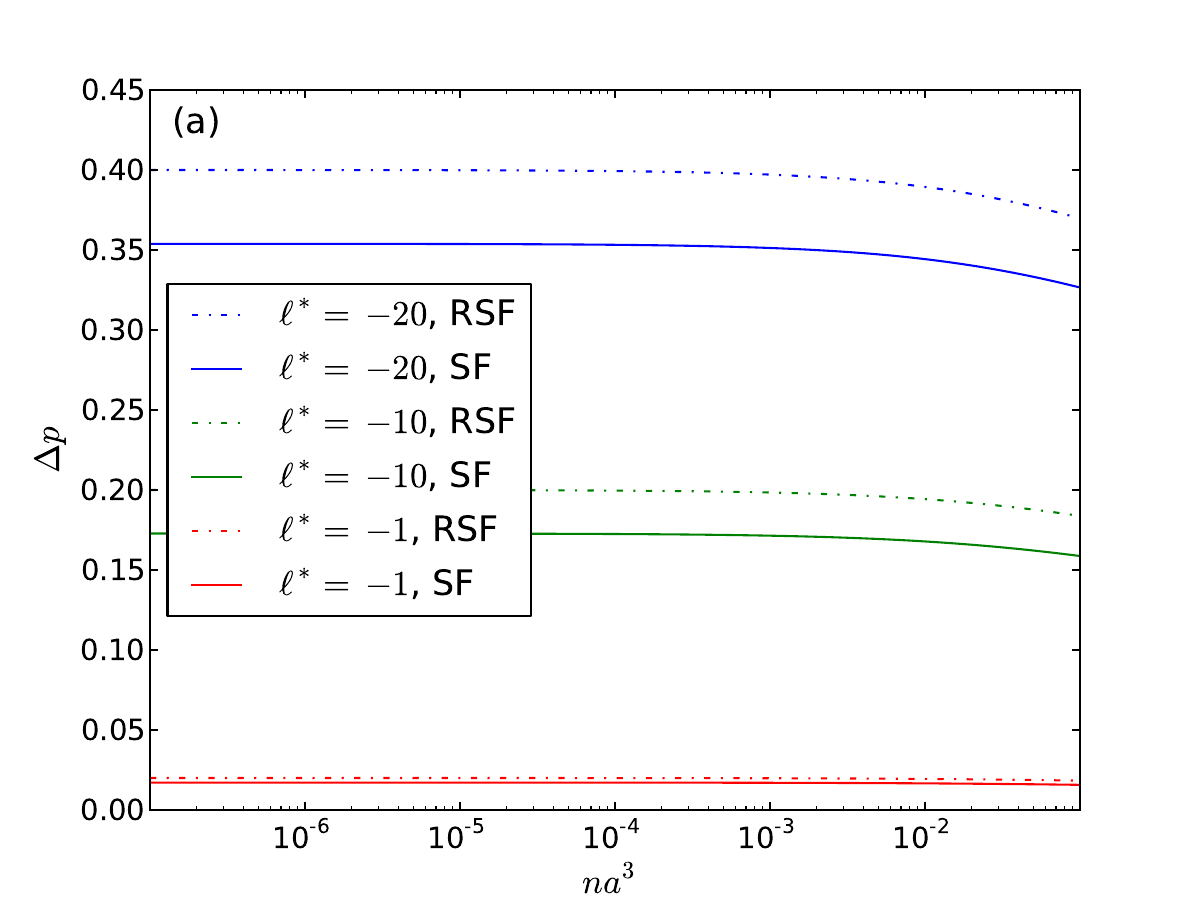}
  \includegraphics[scale=0.4]{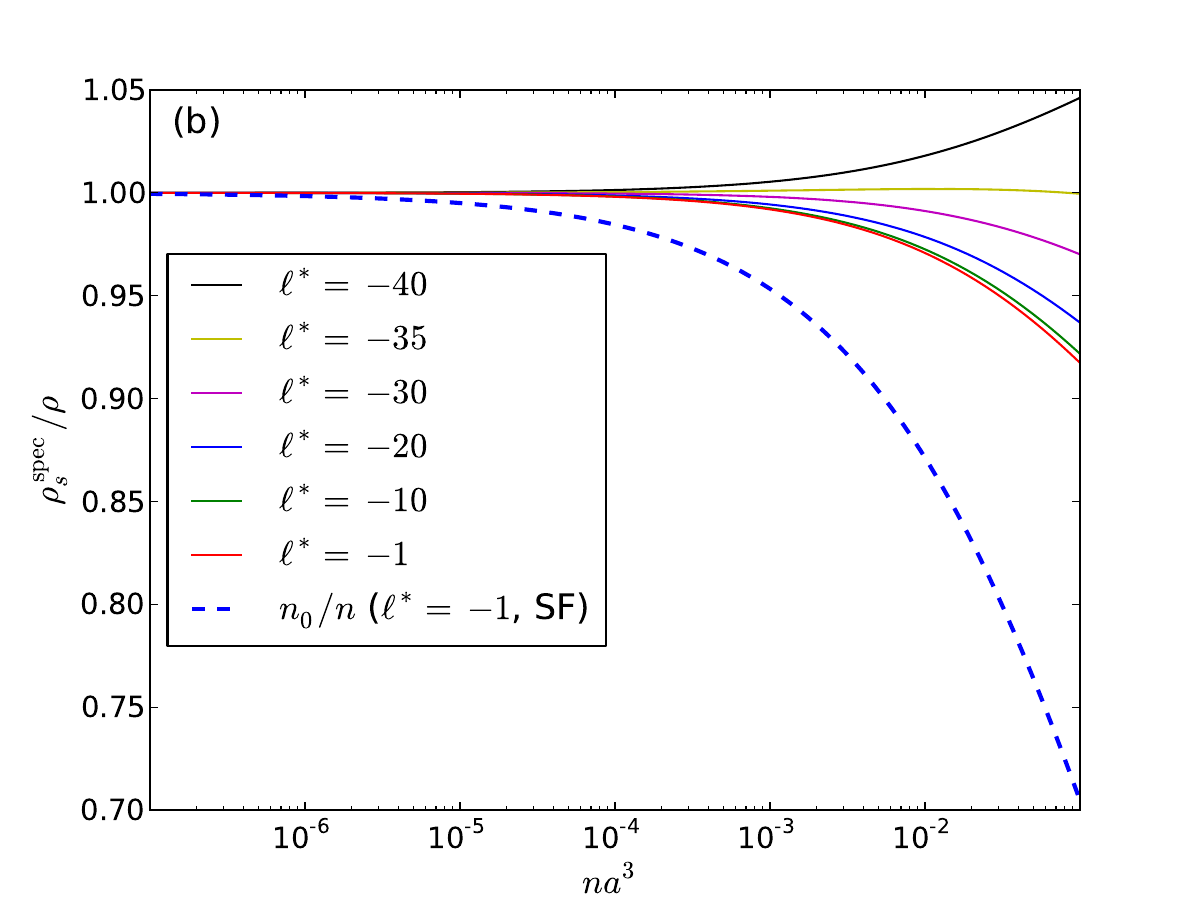}
  \caption{(Color online) Quantitative analysis of
    interaction effects at zero temperature: (a) raw $\Delta p$
    signal; (b) spectroscopically determined superfluid fraction $\rho_s^{\rm spec}/\rho$, and comparison with the condensate fraction
    $n_0/n$. Note that the condensate depletion is significantly
    higher than the deviation of $\rho_s^{\rm spec}/\rho$ from 1. The
    curves are labelled from top to bottom.}
  \label{fig:na3}
\end{figure}

To analyse the quantitative accuracy of this method, we at first keep
$T=0$, \ie we consider the accuracy of the method for the case of a
perfect superfluid where we know that $\langle\ell\rangle = 0$ (or
$\ell^*$ in the relaxed SF). We wish to show that at an interaction
strength where the BEC is reasonably depleted, \eg $n_\textrm{ex}/n
\sim 0.1$, the $100\%$ superfluidity is still recovered by the
spectroscopic method.  The numerical results 
 are shown in
Fig.~\ref{fig:na3}, as a function of increasing dimensionless
interaction strength $na^3$ while keeping the density $n$ fixed.
Fig.~\ref{fig:na3}~(a) shows the raw $\Delta p$ results for the
relaxed ($\dpRSF$) and metastable ($\dpSF$)
superfluids. Fig.~\ref{fig:na3}~(b) shows the resulting
spectroscopically determined superfluid fraction, from
Eq.~(\ref{eq:new-sf-measurement}), using $na^3=10^{-7}$ to re\-pre\-sent
$na^3\approx 0$.  For comparison, in Fig~\ref{fig:na3}~(b) the dashed
line shows the numerically determined condensate fraction $n_0/n$ for
the SF~\cite{footnote4}.
The departure of the spectroscopic measurement of the superfluid
fraction from $1$ [the solid lines in Fig.~\ref{fig:na3}~(b)] shows
that the error in determining the superfluid fraction by the
spectroscopic method increases with interaction strength.  However,
this error is much smaller than the excited fraction
$n_\text{ex}/n$. Therefore, the spectroscopic method allows one to distinguish clearly between the condensate fraction and the superfluid fraction in 
a strongly interacting BEC.

\subsection{Optimal experimental parameters}
\label{subsec:tradeoff}

In the preceding sections we have shown that the spectroscopic
method is capable of providing both a qualitative signature of
superfluidity and a quantitatively accurate way to measure the
superfluid fraction. We now turn to discuss the experimental
parameters to use to optimise this technique as a quantitative
measurement of superfluid fraction.

As was discussed in Section~\ref{sec:corrections}, there is a trade-off
between the signal size and the quantitative accuracy of the method.
In particular, we showed that it is advantageous to have as large a
value of $\Delta \ell$ as possible, and then to increase $\Omega_R$ as
much as possible to improve accuracy but not so much as to make the
signal $\Delta p$ too small. From these considerations we were led to
choose $\Delta \ell =50$ (from practical considerations of achieving
beams of high angular momentum), and we settled on a compromise value
of $\Omega_R=2\pi\times 100\,\mathrm{kHz}$.  However, there still
remains the question of what is the best detuning $\delta$ (and
therefore value of $\ell^*$) to use. Formally, following
Eq.~(\ref{eq:superfluid-measurement}) we should consider the limit
$\ell^* \to 0$, such that the superfluid remains metastable even very
close to $T_c$. But in this limit the signal becomes very small. What is a reasonable value of $\ell^*$ to use? 

\begin{figure}[tbhp]
  \centering
  \includegraphics[scale=0.4]{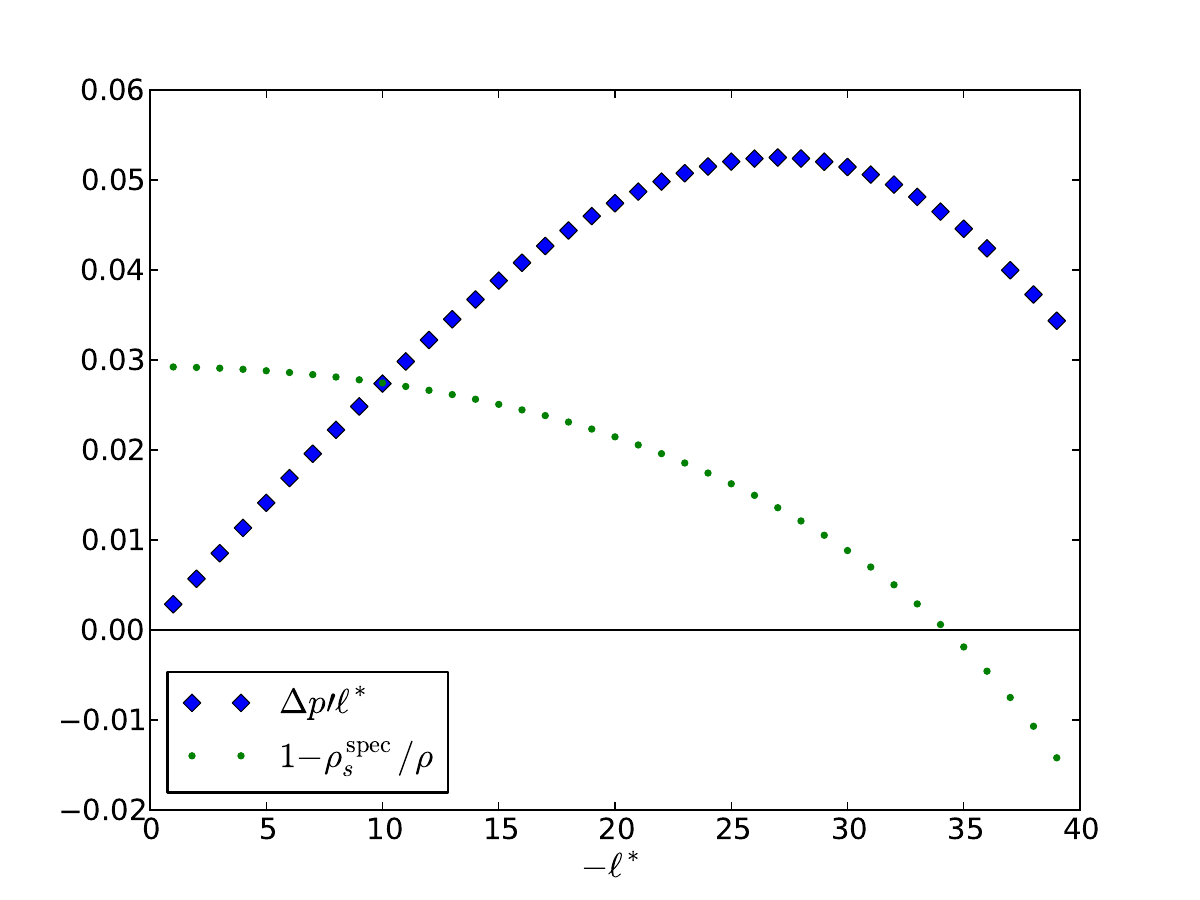}
  \caption{(Color online) Trade-off between signal size and accuracy at zero temperature. Signal size is assumed to be given by $\Delta p' \ell^*$; compare Fig.~\ref{fig:boltzmann}~(c). Accuracy is determined by the deviation of the spectroscopic measurement~\eqref{eq:new-sf-measurement} from the expected unity. This deviation has been calculated at $na^3=10^{-2}$ (corresponding to a condensate depletion $n_\textrm{ex}/n \approx 12\%$), amounting to a vertical slice through Fig.~\ref{fig:na3}~(b) at $na^3=10^{-2}$.}
  \label{fig:signalaccuracy}
\end{figure}

To explore this question, in Fig.~\ref{fig:signalaccuracy} we
present the results of calculations at zero temperature and
$na^3=10^{-2}$, corresponding to a condensate depletion of about $12\%$.

In Fig.~\ref{fig:signalaccuracy} the diamonds show the signal
strength, as given by $\Delta p'\ell^*$. This is the \emph{ideal}
difference between the signals of a normal fluid and a superfluid at
$T=0$, neglecting higher-order corrections to $\Delta p_\ell$. As in
Fig.~\ref{fig:boltzmann}~(c), this is a good approximation to the
actual signal.  These results show that, for the parameters chosen,
the fractional imbalance $\Delta p$ must be measured to an accuracy of
about $0.01$ to $0.05$. While this is a small fractional imbalance, it
should be feasible to detect signals of this size by averaging over
many shots.

The circles in Fig.~\ref{fig:signalaccuracy} display the inaccuracy
in the spectroscopic measurement of the superfluid fraction, \ie 
the relative deviation of $\rho_s^{\rm spec}/\rho$ from the expected value of $1$ at $T=0$.
Hence the circles in
Fig.~\ref{fig:signalaccuracy} correspond to a vertical
slice through Fig.~\ref{fig:na3}~(b) at $na^3=10^{-2}$.  Here, an
inaccuracy of $0.03$  means that a $100\%$ superfluid would be
misinterpreted as only having $97\%$ superfluid fraction.  In
contrast, recall that the condensate fraction is about $88\%$. So this
still allows a clear distinction between superfluid and condensate
fraction.

\begin{figure}[tbhp]
  \centering
  \includegraphics[scale=0.4]{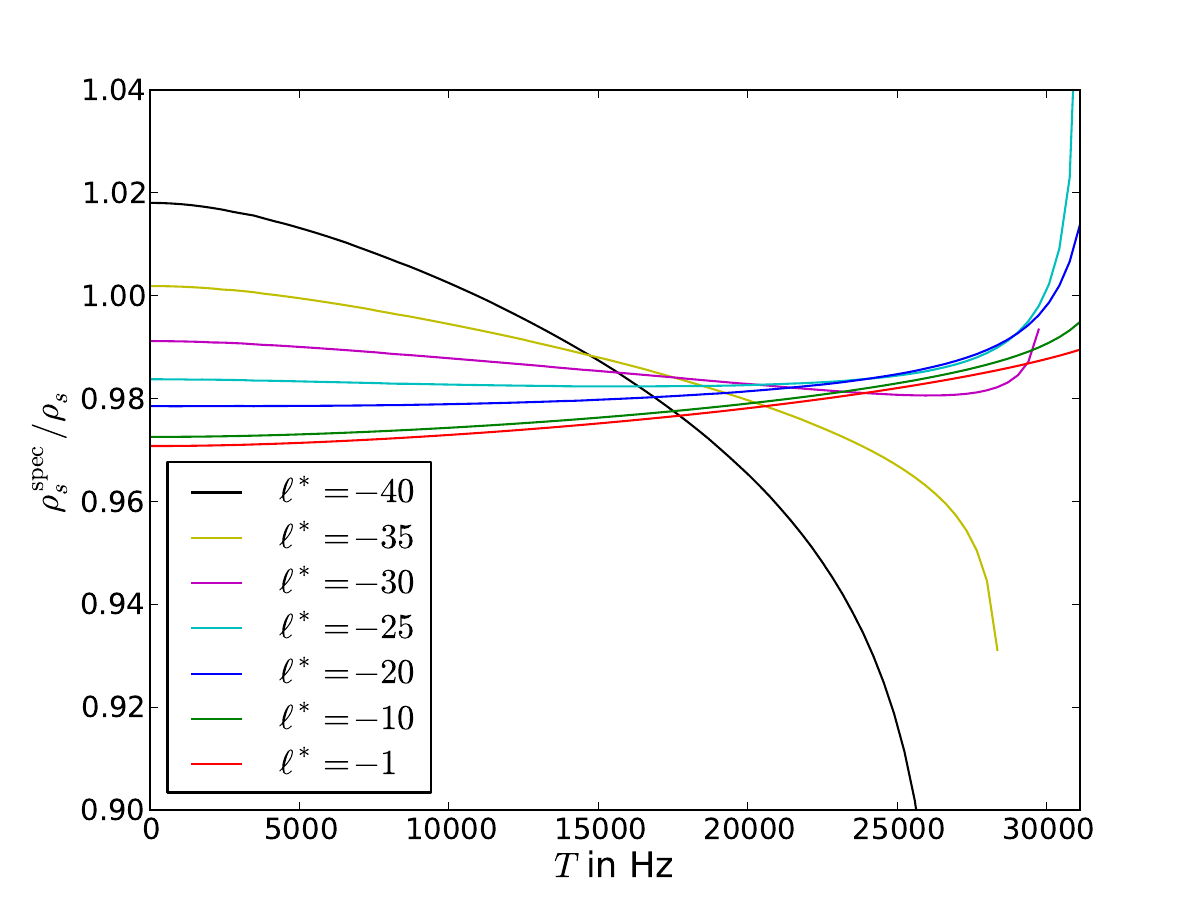}
  \caption{(Color online) Comparison of $\rho_s^{\rm spec}/\rho$ and $\rho_s/\rho = 1 - \langle \ell\rangle/\ell^*$
 at $na^3=10^{-2}$ as a function of temperature (in terms of the \emph{ideal} gas transition temperature $T_c\approx 31.1\,\mathrm{kHz}$). As in Fig.~\ref{fig:hysteresis}, the curves break off for $|\ell^*| > \ell^*_\textrm{crit} \simeq \sqrt{4\pi R^2 a n_0}$. The curves are labelled from top to bottom at the left-hand edge of the graph. Each curve starts at the $y$-value corresponding to the one in Fig.~\ref{fig:na3}~(b) at $na^3=10^{-2}$.}
  \label{fig:with-temperature}
\end{figure}

The results in Fig.~\ref{fig:signalaccuracy} would suggest picking
$|\ell^*| \approx 35$, where the inaccuracy passes through
zero. However, one should note that these results only show the
inaccuracy at zero temperature.
In Fig.~\ref{fig:with-temperature} we plot the ratio $\rho^{\rm
  spec}_s/\rho_s$ as a function of temperature.
Here $\rho_s^{\rm spec}$ is  the spectroscopically measured superfluid
density and
$\rho_s$  is the expected superfluid density,
 as computed from the usual
definition of superfluidity (\ref{eq:superfluid-definition}),
$\rho_s/\rho  = 1 - \langle \ell\rangle/\ell^*$.
The ratio stays close to one over a wide
temperature range, showing that the
spectroscopic method provides a good measure of the superfluid
fraction.

Combining the issue of signal size and accuracy over a range of
temperatures, we find that (for the parameters studied), a good choice
is $|\ell^*|$ between $20$ and $30$. In this range, the signal size is
$\simeq 0.05$; the spectroscopically measured superfluid fraction is
accurate to about $0.02$ (much less than the depleted fraction of
$0.12$); the accuracy remains at this level up to $\simeq 0.9 T_{c}$.

\section{Summary}
\label{sec:summary}

In summary, we have assessed the feasibility of a recent
proposal~\cite{CooperHadzibabic2010} to measure the superfluid fraction
of atomic BECs. This proposal involves the use of optically induced
gauge fields to simulate rotation, and allows the superfluid response
to appear in a spectroscopic signal.  We have calculated the expected
spectroscopic signatures for three-dimensional BECs with uniform
density. One conclusion of our studies is the demonstration that this
technique can be used to obtain a {\it qualitative} experimental
signature of the superfluid response (by comparing the spectroscopic
signals when the order of turning on the gauge field and cooling the
gas is reversed), with a spectroscopic signal that is large
enough to be detectable in experiment. Furthermore, and most
importantly, we have shown that the technique can be used to extract a
{\it quantitative} measurement of the superfluid density. The accuracy
of the measurement technique can be improved at the expense of the
size of the signal.  Using realistic values for the experimental
parameters, we have shown that a compromise can be reached where both
(i)~the signal is sufficiently large to be feasible to measure and
(ii)~the superfluid density is determined to sufficient accuracy to
allow quantitatively useful information to be extracted. Notably, our
results show that the technique can allow a clear experimental
measurement of the distinction between condensate and superfluid
fraction of a strongly interacting Bose gas.  Our results support the
usefulness of this technique~\cite{CooperHadzibabic2010} for measuring
the superfluid fraction, and provide guidance for the parameters
required in future experimental implementations.

\vskip0.1cm

\acknowledgments{We are grateful to Jean Dalibard for helpful comments
  on the manuscript. This work was supported by EPSRC Grant No.\
  EP/I010580/1.}



\begin{thebibliography}{10}
\bibitem{footnote1}
{F}or simplicity, we neglect any quadratic Zeeman effect, assuming that the
  magnetic field is small.
\bibitem{footnotet} {F}or easy comparison with other parameters, for example the Rabi frequency, we express temperatures in frequency units, such that $1\,\mathrm{kHz} = 48\,\mathrm{nK}$.
\bibitem{footnote3}
{S}ince we assumed that the condensate is in $\vecb{p}_c=0$, the last term of
  $\gamma(\vecb{p})$ in~(\protect\ref{eq:bogoliubov-parameters}) refers to the
  condensate state; in general the argument to $\epsilon^0$ has to be changed
  appropriately: $\gamma(\vecb{p}) = \frac{1}{2}
  [\epsilon^0(\vecb{p}_c+\vecb{p})+\epsilon^0(\vecb{p}_c-\vecb{p})] -
  \epsilon^0(\vecb{p}_c)$.
\bibitem{footnote4}
{F}or the parameters used, this is indistinguishable from the condensate
  density determined by self-consistently solving
  Eq.~(\protect\ref{eq:n0-popov}) for $n_0$, showing that non-parabolic
  corrections are irrelevant when considering the condensate fraction.
\end{thebibliography}
\end{document}